\documentclass[a4paper,11pt]{article}
\pdfoutput=1 

\usepackage{jcappub} 

\usepackage{appendix}

\title{Neutrinos and dark energy after Planck and BICEP2: data consistency tests and cosmological parameter constraints}

\author[a]{Jing-Fei Zhang,}
\author[a]{Jia-Jia Geng,}
\author[a,b,1]{Xin Zhang\note{Corresponding author.}}

\affiliation[a]{Department of Physics, College of Sciences, Northeastern University, \\Shenyang
110004, China}
\affiliation[b]{Center for High Energy Physics, Peking University, \\Beijing 100080, China}

\emailAdd{jfzhang@mail.neu.edu.cn}
\emailAdd{gengjiajia163@163.com}
\emailAdd{zhangxin@mail.neu.edu.cn}

\abstract{The detection of the B-mode polarization of the cosmic microwave background (CMB) by the BICEP2 experiment 
implies that the tensor-to-scalar ratio $r$ should be involved in the base standard cosmology. In this paper, we extend the 
$\Lambda$CDM+$r$+neutrino/dark radiation models by replacing the cosmological constant with the dynamical dark energy 
with constant $w$. Four neutrino plus dark energy models are considered, i.e., the $w$CDM+$r$+$\sum m_\nu$, 
$w$CDM+$r$+$N_{\rm eff}$, $w$CDM+$r$+$\sum m_\nu$+$N_{\rm eff}$, and 
$w$CDM+$r$+$N_{\rm eff}$+$m_{\nu,{\rm sterile}}^{\rm eff}$ models. The current observational data considered in this paper 
include the Planck temperature data, the WMAP 9-year polarization 
data, the baryon acoustic oscillation data, the Hubble constant direct measurement data, the Planck 
Sunyaev-Zeldovich cluster counts data, the Planck CMB lensing data, the cosmic shear data, and the 
BICEP2 polarization data. We test the data consistency in the four cosmological models, and then combine the 
consistent data sets to perform joint constraints on the models. 
We focus on the constraints on the parameters $w$, $\sum m_\nu$, $N_{\rm eff}$, and $m_{\nu,{\rm sterile}}^{\rm eff}$.}

\begin{document}
\maketitle
\flushbottom

\section{Introduction}
\label{intro}

Discovery of neutrino oscillations has indicated that neutrinos have nonzero masses. 
Neutrino oscillation experiments can only place limits on the squared mass differences between the 
neutrino mass eigenstates: the solar and reactor experiments observed $\Delta m_{21}^2\simeq 8\times 10^{-5}$ 
eV$^2$, and the atmospheric and accelerator beam experiments observed $\Delta m_{32}^2\simeq 3\times 10^{-3}$ 
eV$^2$. To measure the absolute masses of neutrinos, different experiments are needed. 
In fact, cosmological data have been providing tight limits on the total mass of neutrinos. 

The cosmic microwave background (CMB) observations have been used to constrain the neutrino mass and 
possibly the extra relativistic degrees of freedom (sometimes referred to as ``dark radiation'')~\cite{wmap5,wmap7,wmap9}. 
The latest CMB temperature power spectrum measured by the Planck satellite mission provided the tight limits on 
the total mass of active neutrinos, $\sum m_\nu$, and the  effective number of relativistic species, $N_{\rm eff}$~\cite{planck}. 
For example, the Planck+WP+highL data combination (here, WP denotes the WMAP 9-year polarization data, and  
highL denotes the ACT and SPT temperature data) gives the 95\% confidence level (CL) limits: 
$\sum m_\nu<0.66$ eV for the case of no extra relics ($N_{\rm eff}=3.046$) and $N_{\rm eff}=3.36^{+0.68}_{-0.64}$ for 
the case of minimal-mass normal hierarchy for the neutrino masses (only one massive eigenstate with $m_\nu=0.06$ eV).
Late-time geometric measurements can be used to help reduce some geometric degeneracies and thus improve constraints. 
Therefore, the baryon acoustic oscillation (BAO) data are very useful in the parameter estimation. 
Note also that the BAO data are proven to be in good agreement with the Planck data, so one can always combine CMB data 
with BAO data without any question. The Planck+WP+highL+BAO data combination changes the above limits to: 
$\sum m_\nu<0.23$ eV for the case of no extra relics and $N_{\rm eff}=3.30^{+0.54}_{-0.51}$ for the case of minimal-mass 
normal hierarchy model.

The measurements of the growth of large-scale structure also play a crucial role in constraining the neutrino mass. 
The Planck Collaboration reported a result of counts of rich clusters of galaxies by analyzing its sample of thermal Sunyaev-Zeldovich (SZ) 
clusters, $\sigma_8(\Omega_m/0.27)^{0.3}=0.782\pm 0.010$~\cite{SZ}. And the weak lensing data can also give another combination of 
$\sigma_8$ and $\Omega_m$. The cosmic shear measurement provided by the CFHTLenS survey gave $\sigma_8(\Omega_m/0.27)^{0.46}=0.774\pm 0.040$~\cite{shear}.
However, based on the six-parameter base $\Lambda$CDM cosmology, the Planck data predict more clusters than these astrophysical measurements observe.
The Planck+WP+highL data combination gives $\sigma_8(\Omega_m/0.27)^{0.3}=0.87\pm 0.02$ and $\sigma_8(\Omega_m/0.27)^{0.46}=0.89\pm 0.03$, 
in tensions with the above two measurements at the 4.0$\sigma$ and 2.3$\sigma$ levels, respectively.
In fact, the tensions might just imply non-zero neutrino masses (see, e.g., Ref.~\cite{snu3}). 
Increasing neutrino masses could suppress the growth of structure below their free-streaming length, allowing $\sigma_8$ to be substantially lower.

On the other hand, the measurement of the Hubble constant $H_0$ is also very useful in breaking parameter degeneracies in cosmological constraints, 
but the direct measurement of $H_0$ is in tension with the Planck result based on the standard cosmology. 
The direct measurement gives $H_0=73.8\pm 2.4$ km s$^{-1}$ Mpc$^{-1}$~\cite{h0}, but the six-parameter $\Lambda$CDM model fitting to Planck+WP+highL data  
gives $H_0=67.3\pm 1.2$ km s$^{-1}$ Mpc$^{-1}$; there is a tension at the 2.4$\sigma$ level between the two.
One way of reducing the tension is to increase $N_{\rm eff}$~\cite{planck}. 
Increasing the radiation density at fixed $\theta_\ast$ (to preserve the angular scales of the acoustic peaks) and fixed $z_{\rm eq}$ (to preserve the 
early-ISW effect and so first-peak height) leads to the increase of the Hubble expansion rate before recombination and thus the decrease of the 
age of the universe. The angular scale of the photon diffusion length, $\theta_{\rm D}$, is thus increased, thereby reducing the power in the CMB damping tail 
at a given multipole. The result is that $N_{\rm eff}$ is positively correlated with $H_0$.

Therefore, increasing both $\sum m_\nu$ and $N_{\rm eff}$ could reconcile all the mentioned tensions between Planck and other astrophysical observations, and 
on the other hand the combination of these consistent data sets could determine these parameters more accurately. 
But in this case one has to assume the model in which the massive active neutrinos coexist with some extra radiation degrees of freedom. 
Actually, a more natural choice is to consider the model with light sterile neutrinos. 

The existence of light massive sterile neutrinos is hinted by the anomalies of short-baseline neutrino experiments~\cite{lsnd,miniboone,reactor,Giunti:2012tn,Giunti:2013aea}. 
There is evidence for oscillations at a $\Delta m^2$ of about 1 eV$^2$ from these experiments.
Thus, additional neutrino masses are required. And the additional types of neutrino have to be sterile neutrinos so that they 
do not interact by the weak interaction and do not affect the width of $Z^0$.
Through their mixing with active neutrinos and their gravitational interactions, the sterile neutrinos could have a considerable 
effect on astrophysics and cosmology.
Therefore, it is particularly important to search for cosmological evidence of sterile neutrinos through their gravitational effects on 
the large-scale structure formation and the evolution of the universe. 
When considering sterile neutrinos in a cosmological model, two extra parameters, $N_{\rm eff}$ and $m_{\nu,{\rm sterile}}^{\rm eff}$, 
are added. Hence, in a cosmological model with sterile neutrinos, the tensions of Planck with $H_0$ direct measurement, counts of Planck 
SZ clusters, and cosmic shear can be substantially relieved~\cite{snu1,snu2,snu3}.

Recently, the BICEP2 Collaboration reported the detection of the B-mode polarization of the CMB~\cite{bicep2}. 
If the treatment of the foreground model is correct, the BICEP2's result indicates the discovery of the primordial gravitational waves (PGWs). 
The frontiers of fundamental physics will be pushed forward in an unprecedented way as long as the BICEP2's result is confirmed by upcoming experiments. 
Adopting the $\Lambda$CDM+$r$ model, the fit to the observed B-mode power spectrum gives an unexpectedly large tensor-to-scalar ratio, $r=0.20^{+0.07}_{-0.05}$, 
with $r=0$ disfavored at the 7$\sigma$ level~\cite{bicep2}.
This result is in tension with the upper limit, $r<0.11$ (95\% CL), given by the fit to the combined Planck+WP+highL data~\cite{planck}. 
To explain and/or reduce this tension, numerous proposals have been put forward; see, 
e.g., Refs.~\cite{Liu:2014mpa,Harigaya:2014qza,Nakayama:2014koa,Brandenberger:2014faa,
Contaldi:2014zua,Miranda:2014wga,Gerbino:2014eqa,McDonald:2014kia,Hazra:2014a,Hazra:2014b,Kehagias:2014wza,Lyth:2014yya,Bonvin:2014xia,Lizarraga:2014eaa,Moss:2014cra,Chluba:2014uba}.
An intriguing mechanism for relieving this tension is to involve a sterile neutrino species in the cosmological model with PGWs~\cite{zx14,WHu14,sterile2,sterile3,nus14a,nus14b}. 
In Ref.~\cite{zx14}, the model with PGWs and sterile neutrinos is called $\Lambda$CDM+$r$+$\nu_s$ model (or $\Lambda$CDM+$r$+$N_{\rm eff}$+$m_{\nu,{\rm sterile}}^{\rm eff}$ model).
It was shown in Refs.~\cite{zx14,WHu14} that in the $\Lambda$CDM+$r$+$\nu_s$ model the tension between Planck and BICEP2 is well relieved, and meanwhile, 
the other tensions of Planck with other astrophysical observations, such as the $H_0$ direct measurement, the cluster counts, and the cosmic shear measurement, can all be 
significantly reduced. In addition, the current consistent cosmological data could provide independent evidence for the existence of sterile neutrino at high statistical significance, 
i.e., a joint analysis of current data prefer $\Delta N_{\rm eff}\equiv N_{\rm eff}-3.046>0$ at the 2.7$\sigma$ level and a nonzero mass of sterile neutrino at the 3.9$\sigma$ level~\cite{zx14}. 
In Ref.~\cite{sterile2}, other typical models of neutrino and dark radiation were analyzed with the current observational data in detail.

In this paper, we investigate the effects of dynamical dark energy on the fits of models of neutrinos and dark radiation to the current observations.
In Ref.~\cite{hde}, it was demonstrated that the tension between Planck and $H_0$ direct measurement might hint that dark energy is not the cosmological constant; 
once a dynamical dark energy is adopted, the tension of Planck with $H_0$ measurement could be well relieved.
Also, it is known that due to the repulsive gravitational force, dark energy could suppress the growth of large-scale structure. 
Therefore, we wish to see if dark energy together with neutrinos could further reduce the tensions between Planck and other astrophysical observations.
On the other hand, we will combine the consistent data sets to constrain the cosmological parameters concerning dark energy and neutrinos. 

The paper is organized as follows. In Sec.~\ref{method}, we briefly present the cosmological models and the observational data we use in this work.
In Sec.~\ref{result}, we test the data consistency for the models and present the fit results. Conclusion is given in Sec.~\ref{concl}.
More detailed fit results are given in Appendix~\ref{aa}.

\section{Methodology}
\label{method}

In this paper, we consider the extension of the $\Lambda$CDM cosmology to models in which dark energy has a constant $w$, i.e., the $w$CDM cosmology.
Of course, since the PGWs are likely to have been detected by BICEP2, the basic framework should be the $w$CDM+$r$ model. 
Further extensions considered in this paper include: 
(i) active neutrinos with additional parameter $\sum m_\nu$, 
(ii) extra dark radiation with additional parameter $N_{\rm eff}$, 
(iii) active neutrinos coexisting with extra relativistic degrees of freedom, with additional parameters $\sum m_\nu$ and $N_{\rm eff}$, 
and (iv) massive sterile neutrinos with additional parameters $N_{\rm eff}$ and $m_{\nu, {\rm sterile}}^{\rm eff}$.

The conventions used in this paper are consistent with those adopted by the Planck team \cite{planck}, i.e., those used in the {\tt camb} Boltzmann code. 
The base parameters for the basic eight-parameter $w$CDM+$r$ model are:
$$\{\omega_b,~\omega_c,~100\theta_{\rm MC},~\tau,~w,~n_s,~\ln (10^{10}A_s),~r_{0.05}\},$$
where $\omega_b\equiv \Omega_b h^2$ and $\omega_c\equiv \Omega_c h^2$ are the baryon and cold dark matter densities today, respectively, 
$\theta_{\rm MC}$ is the approximation used in {\tt CosmoMC} to $r_s(z_\ast)/D_A(z_\ast)$ (the angular size of the sound horizon at the time of last-scattering), 
$\tau$ is the Thomson scattering optical depth due to reionization, 
$w$ is the dark energy equation of state parameter, $n_s$ and $A_s$ are the power-law spectral index and power of the 
primordial curvature perturbations, respectively, and $r_{0.05}$ is the tensor-to-scalar ratio at the pivot scale $k_0=0.05$ Mpc$^{-1}$.
Flat priors for the base parameters are used. Note also that the prior ranges for the base parameters are chosen to be much wider than the posterior 
in order not to affect the results of parameter estimation. 
We use the {\tt CosmoMC} package~\cite{Lewis:2002ah} to infer the posterior probability distributions of parameters. 

Next, we describe the observational data sets used in this paper. 
Actually, this work is an extension of our previous works \cite{zx14,sterile2} to the models of dark energy with constant $w$, thus we will use the same data sets to 
Refs.~\cite{zx14,sterile2} in order to be easier for a direct comparison. 
The data sets we use include the CMB (Planck+WP), BAO, $H_0$, SZ, Lensing (CMB lensing + weak lensing), and BICEP2.

{\bf Planck+WP.} We use the CMB TT angular power spectrum data from the first release of Planck \cite{planck}, combined with the CMB large-scale TE and EE polarization power 
spectrum data form the 9-yr release of WMAP \cite{wmap9}. Note that in some occasions we also abbreviate Planck+WP to CMB for convenience. 

{\bf BAO.} We use the latest BAO measurements from the data release (DR) 11 of the Baryon Oscillation Spectroscopic Survey (BOSS) (part of SDSS-III): 
$D_V(0.32)(r_{d,{\rm fid}}/r_d)=(1264\pm 25)$~Mpc and $D_V(0.57)(r_{d,{\rm fid}}/r_d)=(2056\pm 20)$~Mpc, with 
$r_{d,{\rm fid}}=149.28$~Mpc~\cite{boss}. 
Note also that there are some other BAO data sets, e.g., 6dFGS ($z=0.1$)~\cite{6df}, SDSS-DR7 ($z=0.35$)~\cite{sdss7}, and 
WiggleZ ($z=0.44$, 0.60, and 0.73)~\cite{wigglez}, where 
the three data from the WiggleZ survey are correlated (with the inverse covariance matrix given by Ref.~\cite{wigglez}). 
This work is in accordance with our previous papers \cite{zx14,sterile2} for the use of the BAO data, i.e.,  we only use the latest two most accurate BAO data from the BOSS-DR11. 
This is sufficient for our purpose in breaking the CMB parameter degeneracies. 

$\boldsymbol{H_0.}$ We use the direct measurement of the Hubble constant from the Hubble Space Telescope (HST) observations, 
$H_0=(73.8\pm 2.4)~{\rm km}~{\rm s}^{-1}~{\rm Mpc}^{-1}$~\cite{h0}.

{\bf SZ.} We use the result of the counts of clusters of galaxies from the sample of Planck SZ clusters, 
$\sigma_8(\Omega_m/0.27)^{0.3}=0.782\pm 0.010$~\cite{SZ}. 
Note that this result is derived by using the mass function given by Tinker et al.  \cite{Tinker2008}; a different mass function given by Watson et al. \cite{Watson2013} 
leads to a slightly different value, i.e., 
$\sigma_8(\Omega_m/0.27)^{0.3}=0.802\pm 0.014$.
Moreover, the result also depends on the bias $(1-b)$ that is assumed to account for all the possible 
observational biases including departure from hydrostatic equilibrium, absolute instrument calibration, temperature inhomogeneities, 
residual selection bias, etc.
Numerical simulations taking into account of several ingredients of gas physics of clusters give the result of the bias of $(1-b)=0.8^{+0.2}_{-0.1}$. 
Adopting the central value, $(1-b)=0.8$, the Planck team found that the constraints on $\Omega_m$ and $\sigma_8$ are in good agreement with 
previous measurements using clusters of galaxies~\cite{SZ}. The result of $\sigma_8(\Omega_m/0.27)^{0.3}$ quoted in this paper is derived by fixing $(1-b)=0.8$.  
The result is changed to $\sigma_8(\Omega_m/0.27)^{0.3}=0.764\pm 0.025$, if the bias is allowed to vary in the range of $[0.7,1]$. 
In Table 2 of Ref.~\cite{SZ}, other values of $\sigma_8(\Omega_m/0.27)^{0.3}$ from various data combinations and analysis methods are also given. 
But in this paper, in accordance with previous works by us and by other authors, we choose to use the result of $\sigma_8(\Omega_m/0.27)^{0.3}=0.78\pm 0.01$.

{\bf Lensing.} We use two kinds of lensing data, i.e., the CMB lensing power spectrum $C_\ell^{\phi\phi}$ from the Planck mission \cite{cmblensing} and 
the cosmic shear measurement of weak lensing from the CFHTLenS survey, $\sigma_8(\Omega_m/0.27)^{0.46}=0.774\pm 0.040$ \cite{shear}. 
Of course, they are two absolutely different, physically and observationally independent data sets. 

{\bf BICEP2.} We use the CMB angular power spectra (TT, TE, EE, and BB) data from the BICEP2~\cite{bicep2}.

Since it has been proven that the Planck data are in good agreement with the BAO data, we can alway safely combine Planck+WP with BAO. 
So the Planck+WP+BAO combination is the basic data combination used in this paper. 
In the six-parameter base $\Lambda$CDM model, it was shown that the Planck data are in tension with the $H_0$ direct measurement, the cluster counts, 
and the cosmic shear measurement at the 2--3 $\sigma$ level~\cite{planck}. In the seven-parameter $\Lambda$CDM+$r$ model, the Planck temperature data 
are also in tension with the BICEP2 polarization data~\cite{bicep2}. 
It has been demonstrated that the ingredients such as dark energy, dark radiation, and neutrinos could help relieve these tensions 
to some extent~\cite{zx14,WHu14,sterile2,sterile3,nus14a,nus14b,hde}. 
Therefore, in this paper, we will test the data consistency in the cosmological models with dark energy, dark radiation, and neutrinos.

As mentioned above, we consider four models in this paper, i.e., (i) the $w$CDM+$r$+$\sum m_\nu$ model, (ii) the $w$CDM+$r$+$N_{\rm eff}$ model, 
(iii) the $w$CDM+$r$+$\sum m_\nu$+$N_{\rm eff}$ model, and (iv) the $w$CDM+$r$+$N_{\rm eff}$+$m_{\nu,{\rm sterile}}^{\rm eff}$ model. 
We will constrain these models by using the Planck + WP + BAO data, and then compare the derived results with the observations of $H_0$, SZ cluster counts, 
cosmic shear, and tensor-to-scalar ratio. This could test if in these models the Planck+WP+BAO data are consistent with these astrophysical observations. 
If these data sets are consistent, we can then combine these data together and use them to constrain the parameters concerning dark energy, dark radiation, 
neutrinos, and so on. But if the data are still in tension, the direct combination of different data sets is not appropriate. 
The subsequent data fits are based on this data consistency test analysis, and the fit results will be discussed in detail.

\section{Results and discussion}
\label{result}

\subsection{Data consistency tests}

\begin{figure}[tbp]
\centering 
\includegraphics[scale=0.5]{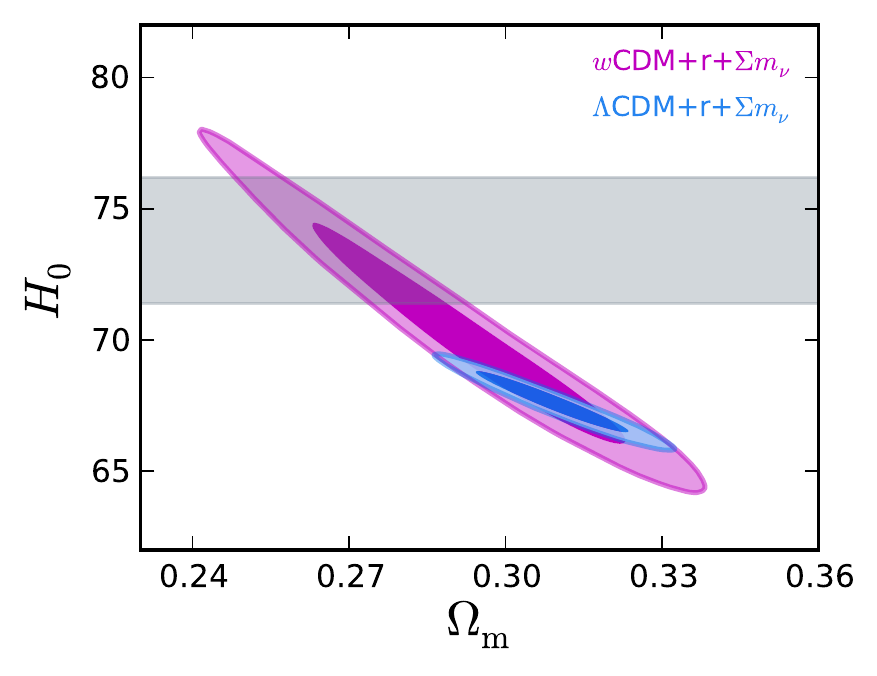}
\includegraphics[scale=0.5]{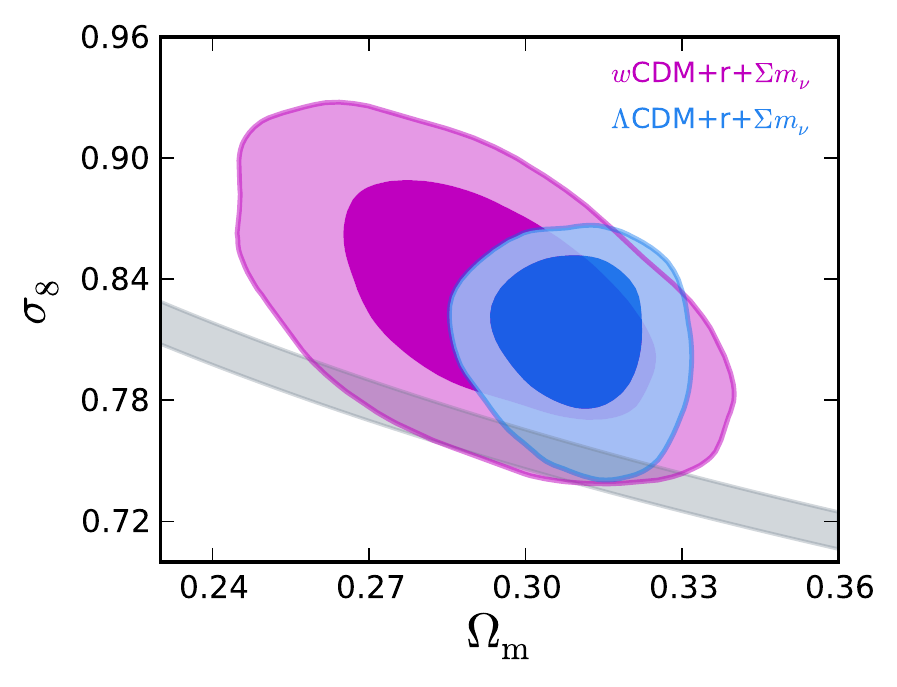}
\includegraphics[scale=0.5]{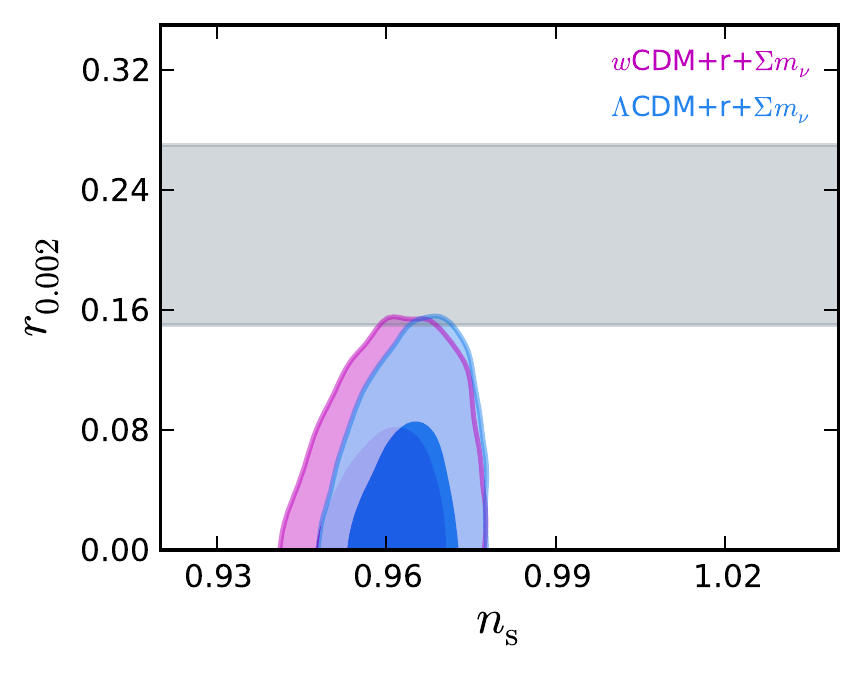}
\includegraphics[scale=0.5]{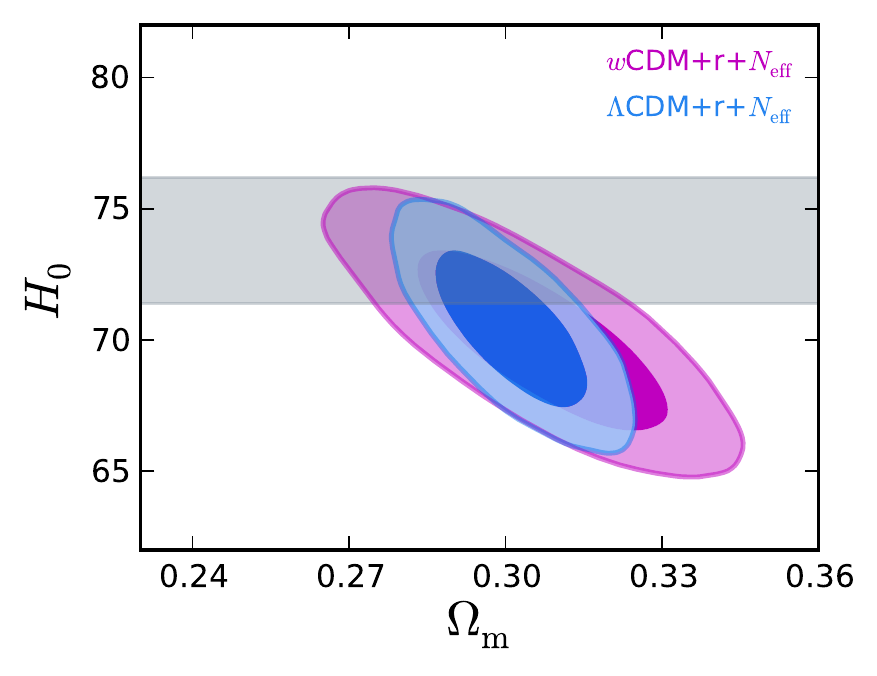}
\includegraphics[scale=0.5]{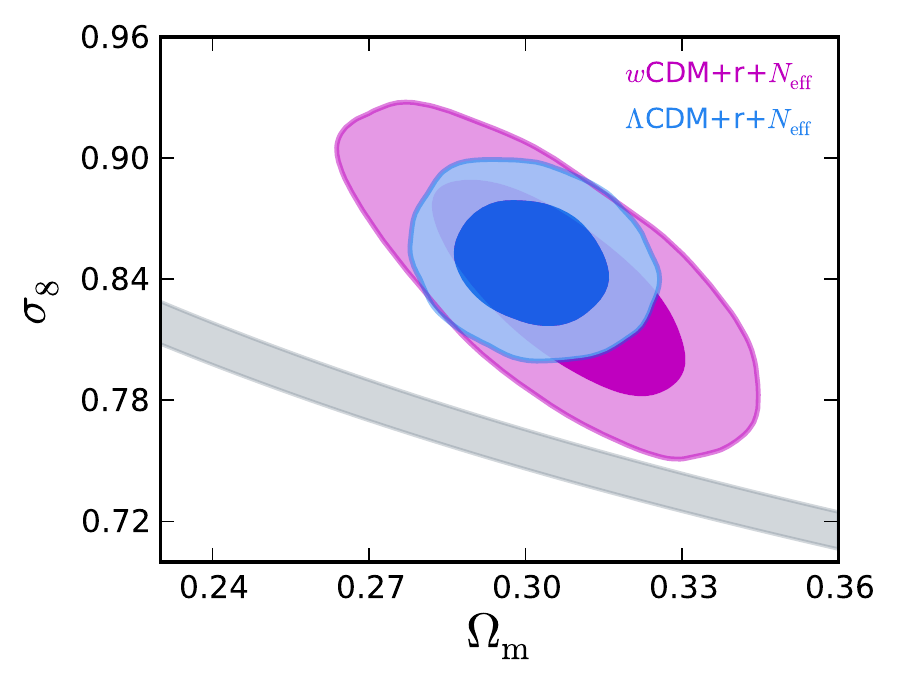}
\includegraphics[scale=0.5]{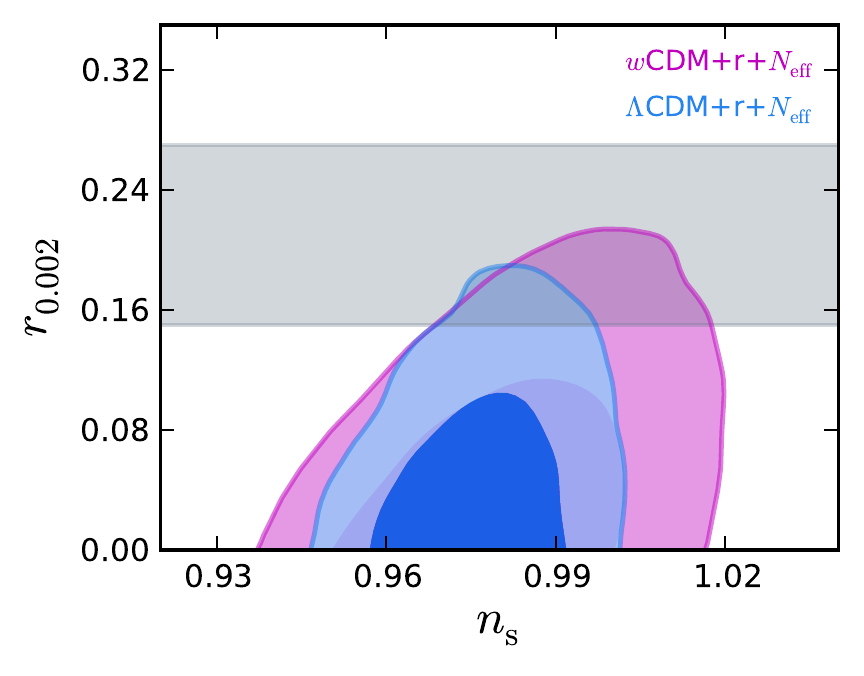}
\includegraphics[scale=0.5]{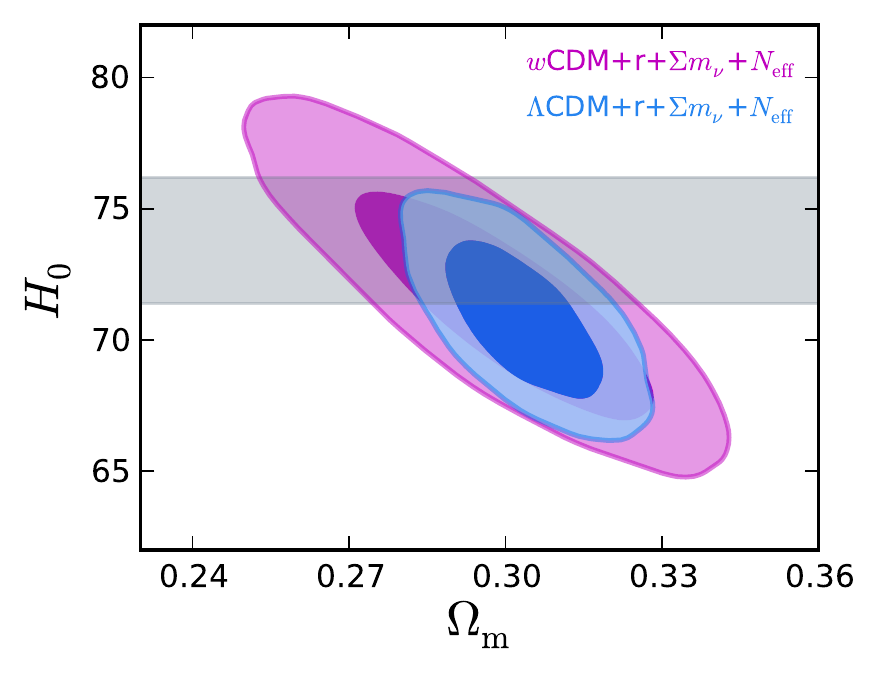}
\includegraphics[scale=0.5]{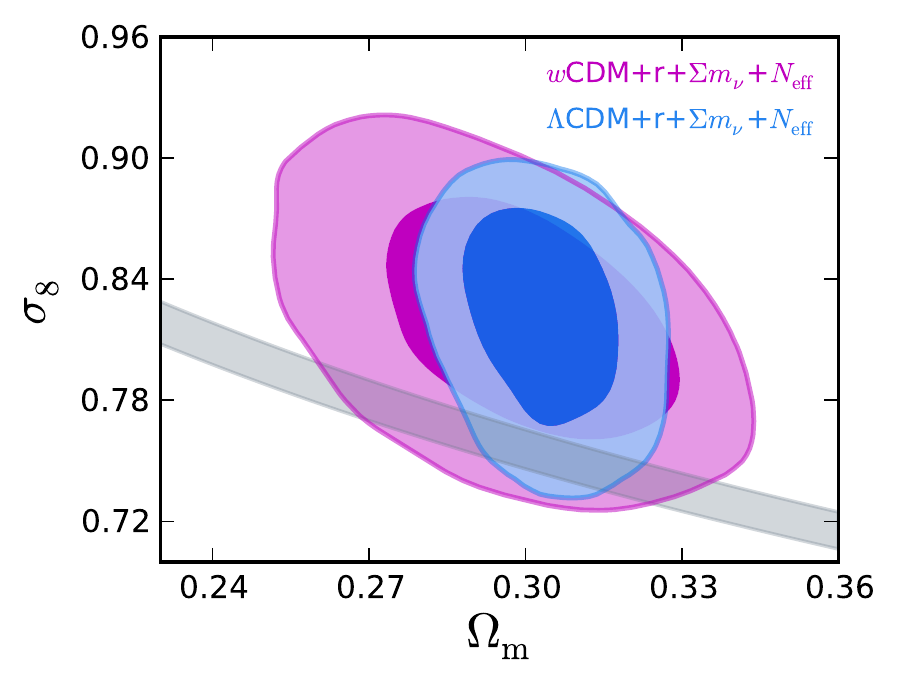}
\includegraphics[scale=0.5]{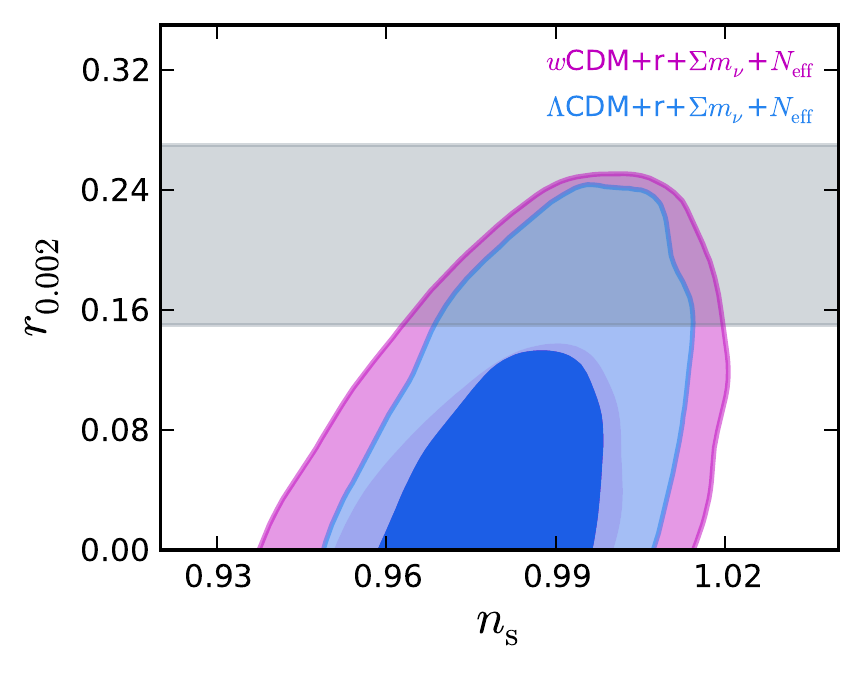}
\includegraphics[scale=0.5]{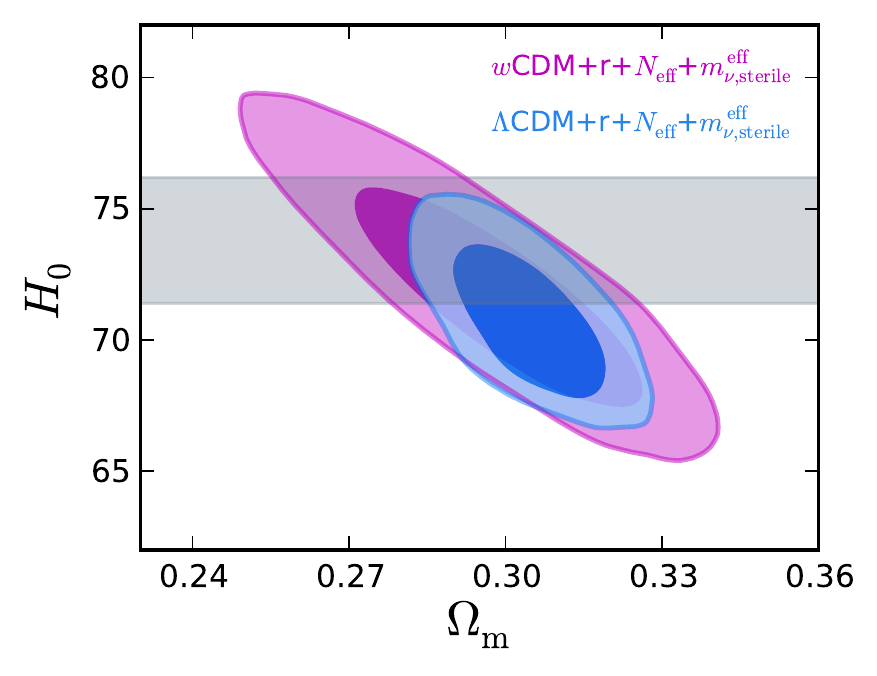}
\includegraphics[scale=0.5]{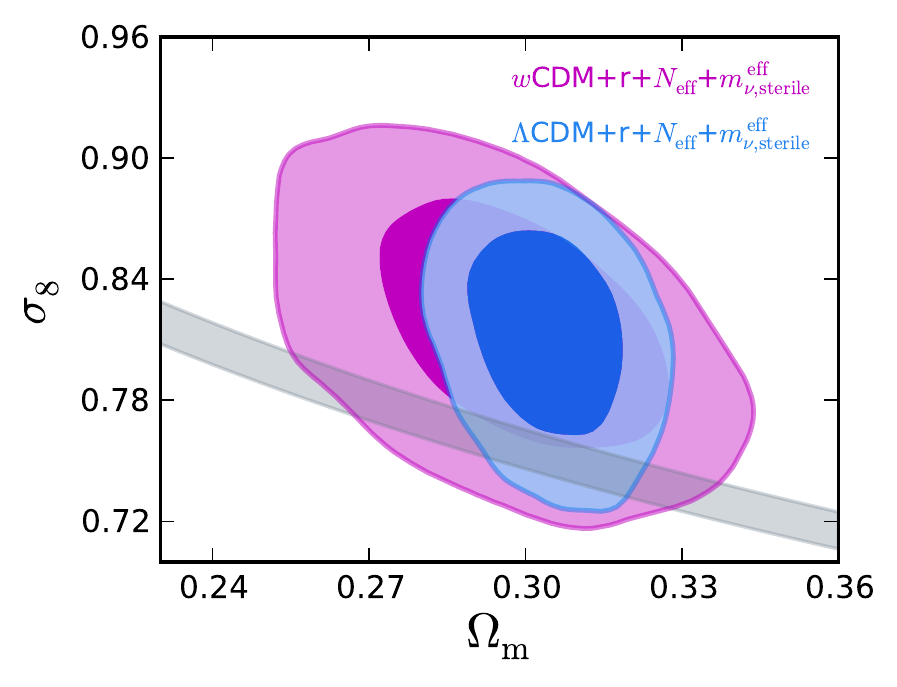}
\includegraphics[scale=0.5]{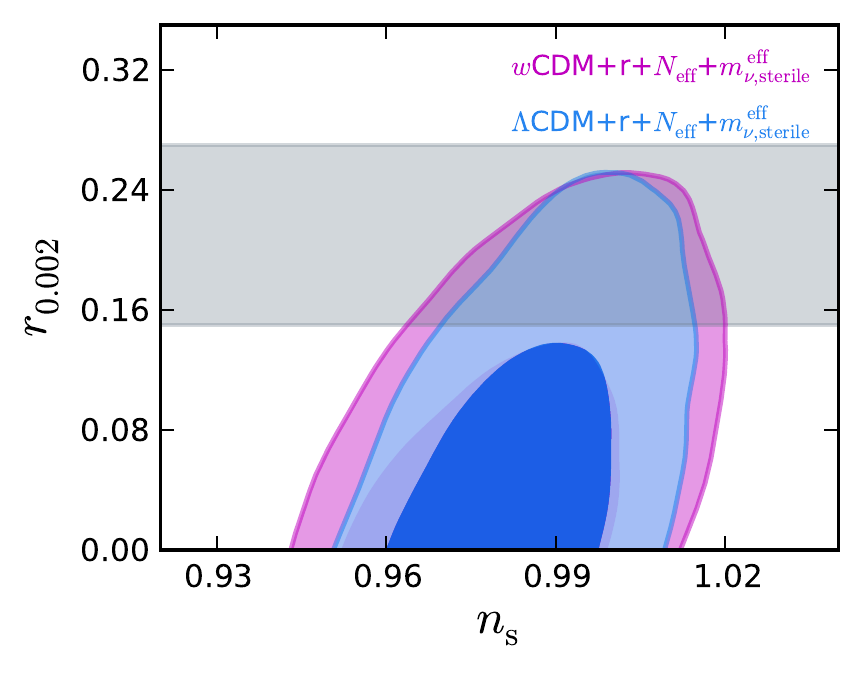}
\hfill
\caption{\label{fig11} The Planck+WP+BAO constraints in the $\Omega_m$--$H_0$, $\Omega_m$--$\sigma_8$, and $n_s$--$r_{0.002}$ planes 
for the $\Lambda$CDM/$w$CDM+$r$+$\sum m_\nu$ models (first-row panels), the $\Lambda$CDM/$w$CDM+$r$+$N_{\rm eff}$ models (second-row panels), 
the $\Lambda$CDM/$w$CDM+$r$+$\sum m_\nu$+$N_{\rm eff}$ models (third-row panels), and the $\Lambda$CDM/$w$CDM+$r$+$N_{\rm eff}$+$m_{\nu,{\rm sterile}}^{\rm eff}$ models 
(fourth-row panels). The gray bands stand for the observational results, i.e., the HST result of Hubble constant $H_0=73.8\pm 2.4$ km s$^{-1}$ Mpc$^{-1}$~\cite{h0}, 
the Planck result of SZ cluster counts $\sigma_8(\Omega_m/0.27)^{0.3}=0.782\pm 0.010$~\cite{SZ}, and 
the BICEP2 result of tensor-to-scalar ratio $r=0.20^{+0.07}_{-0.05}$~\cite{bicep2}.}
\end{figure}

We first constrain the four models using the Planck+WP+BAO data. 
Figure \ref{fig11} shows the fit results in the $\Omega_m$--$H_0$, $\Omega_m$--$\sigma_8$, and $n_s$--$r_{0.002}$ planes.
To exhibit the effects of dynamical dark energy, we also compare the results of $w$CDM + neutrino models with those of $\Lambda$CDM + neutrino 
models (purple contours vs. blue contours).
The first-row panels are for the $\Lambda$CDM/$w$CDM+$r$+$\sum m_\nu$ models, 
the second-row panels are for the $\Lambda$CDM/$w$CDM+$r$+$N_{\rm eff}$ models, 
the third-row panels are for the $\Lambda$CDM/$w$CDM+$r$+$\sum m_\nu$+$N_{\rm eff}$ models, 
and the fourth-row panels are for the $\Lambda$CDM/$w$CDM+$r$+$N_{\rm eff}$+$m_{\nu,{\rm sterile}}^{\rm eff}$ models. 
The gray bands are from the observational results, i.e., the HST result of Hubble constant $H_0=73.8\pm 2.4$ km s$^{-1}$ Mpc$^{-1}$~\cite{h0}, 
the Planck result of SZ cluster counts $\sigma_8(\Omega_m/0.27)^{0.3}=0.782\pm 0.010$~\cite{SZ}, and 
the BICEP2 result of tensor-to-scalar ratio $r=0.20^{+0.07}_{-0.05}$~\cite{bicep2}.
From the comparison of the contours from CMB+BAO  with the bands from other astrophysical observations, one can 
directly test the consistency between different data sets.

First, we discuss the results of the $\Lambda$CDM/$w$CDM+$r$+$\sum m_\nu$ models (see the first-row panels). 
From the $\Omega_m$--$H_0$ plane, we clearly see that the tension on $H_0$  appearing in the $\Lambda$CDM-based model 
is greatly relieved once a dynamical dark energy is considered. But even in the $w$CDM-based model, $\Omega_m$ and $H_0$ are still in strong degeneracy. 
Since $w$ is anti-correlated with $H_0$, we infer that a larger $H_0$ may prefer a phantom energy with $w<-1$. 
From the $\Omega_m$--$\sigma_8$ plane, we see that the consideration of dynamical dark energy amplifies the parameter space but nearly does not 
improve the reduction of the tension in the $\Omega_m$--$\sigma_8$ plane. 
Since $w$ is anti-correlated with $\sigma_8$ but positively correlated with $\Omega_m$, the effect of $w$ on $\sigma_8(\Omega_m/0.27)^{0.3}$ 
would be canceled a lot, leading to the fact that only $\sum m_\nu$ plays a significant role in relieving the tension between Planck CMB and SZ cluster counts. 
From the $n_s$--$r_{0.002}$ plane, we see that $w$ could not affect the value of tensor-to-scalar ratio.

Next, we discuss the results of the $\Lambda$CDM/$w$CDM+$r$+$N_{\rm eff}$ models (see the second-row panels). 
In the base $\Lambda$CDM model, the Planck prediction of $H_0$ is in tension with the direct measurement of $H_0$ at about the 2.5$\sigma$ level, 
but once $N_{\rm eff}$ is considered to be free, this tension could be relieved greatly, as shown by the blue contours in the $\Omega_m$--$H_0$ plane 
for the $\Lambda$CDM+$r$+$N_{\rm eff}$ model. 
Further introducing $w$ does not lead to more impact on $H_0$ (as shown by the purple contours in the $\Omega_m$--$H_0$ plane), 
from which one can infer that the fit of this model to the CMB+BAO data would give the 
result of $w$ around $-1$.  
In the $\Omega_m$--$\sigma_8$ plane, we can see that the parameter $N_{\rm eff}$ does not help reduce the tension between Planck data and cluster data, 
but the parameter $w$ could play a significant role in relieving this tension. 
Due to the repulsive gravitational force of dark energy, the growth of large-scale structure is suppressed by dark energy. 
Contrary to our intuition, larger $w$ gives rise to more suppression of the structure growth; in other words, larger $w$ leads to smaller $\sigma_8$ 
(or, $w$ and $\sigma_8$ are in the anti-correlation). 
The reason is that for larger $w$ and fixed present dark energy density, dark energy comes to dominate earlier, causing the time of suppressing the growth 
of linear matter perturbation to be longer. 
Since $w$ is positively correlated with $\Omega_m$, larger $w$ would lead to larger $\Omega_m$. 
The effects of $w$ on $\sigma_8$ and $\Omega_m$ are contrary, and thus some offset happens for the quantity $\sigma_8(\Omega_m/0.27)^{0.3}$, 
resulting in that increasing $w$ still could not well relieve the tension 
(as shown by the purple contours in the $\Omega_m$--$\sigma_8$ plane). 
Since $N_{\rm eff}$ is positively correlated with $n_s$, and $w$ is also positively correlated with $n_s$, 
they can both contribute to enhance the value of $r$, but it seems that even in the $w$CDM+$r$+$N_{\rm eff}$ model the tension of $r$ is still not well relieved.



\begin{table}[tbp]\tiny
\centering
\setlength\tabcolsep{2.5pt}
\renewcommand{\arraystretch}{1.5}
\begin{tabular}{|lcccccccccccc|}
\hline
 &\multicolumn{2}{c}{$\Lambda$CDM+$r$} & & \multicolumn{2}{c}{$\Lambda$CDM+$r$+$N_{\rm{eff}}$+$m_{\nu,{\rm sterile}}^{\rm eff}$} & & \multicolumn{2}{c}{$w$CDM+$r$} & &\multicolumn{2}{c}{$w$CDM+$r$+$N_{\rm{eff}}$+$m_{\nu,{\rm sterile}}^{\rm eff}$} & \\
\cline{2-3}\cline{5-6}\cline{8-9}\cline{11-12}
Parameter & 68\% limits & tension && 68\% limits & tension && 68\% limits & tension && 68\% limits & tension & \\
\hline
$H_0$&$67.80^{+0.64}_{-0.63}$&$2.4\sigma$&&$70.8^{+1.7}_{-2.1}$&$1.0\sigma$&&$69.0^{+1.8}_{-2.6}$&$1.6\sigma$&&$71.9^{+2.3}_{-3.2}$&$0.6\sigma$&\\
$\sigma_8(\Omega_{\rm{m}}/0.27)^{0.3}$ & $0.857\pm0.015$ &$4.3\sigma$&& $0.842^{+0.038}_{-0.029}$&$2.0\sigma$&&$0.868\pm{0.026}$&$3.2\sigma$&&$0.840^{+0.039}_{-0.034}$&$1.7\sigma$& \\
$\sigma_8(\Omega_{\rm{m}}/0.27)^{0.46}$ & $0.876^{+0.019}_{-0.018}$ &$2.3\sigma$&& $0.858^{+0.038}_{-0.030}$&$1.7\sigma$&&$0.882^{+0.024}_{-0.023}$&$2.3\sigma$&&$0.853^{+0.041}_{-0.033}$&$1.5\sigma$& \\
$r_{0.002}$&$<0.12~(95\%)$&$ $&&$<0.20~(95\%)$&$ $&&$<0.11~(95\%)$&&&$<0.20~(95\%)$&$ $&\\
\hline
\end{tabular}
\caption{\label{tab01} Predictions of the values of $H_0$, $\sigma_8(\Omega_{\rm{m}}/0.27)^{0.3}$, $\sigma_8(\Omega_{\rm{m}}/0.27)^{0.46}$, and $r_{0.002}$
given by the $\Lambda$CDM+$r$, $\Lambda$CDM+$r$+$N_{\rm{eff}}$+$m_{\nu,{\rm sterile}}^{\rm eff}$, $w$CDM+$r$, and $w$CDM+$r$+$N_{\rm{eff}}$+$m_{\nu,{\rm sterile}}^{\rm eff}$
models under the constraints from Planck+WP+BAO. Quantified levels of the tensions with the observational results are also given.}
\end{table}

At last, we discuss the results of both the $\Lambda$CDM/$w$CDM+$r$+$\sum m_\nu$+$N_{\rm eff}$ models and 
the $\Lambda$CDM/$w$CDM+$r$+$N_{\rm eff}$+$m_{\nu,{\rm sterile}}^{\rm eff}$ models (see the third-row and the last-row panels).
Since the results of the two cases are very similar (see also Ref.~\cite{sterile2}), we only focus on the case of sterile neutrino. 
We can see clearly that in the models involving sterile neutrinos all the tensions could be well relieved; see also Refs.~\cite{zx14,WHu14,sterile2,sterile3,nus14a,nus14b}. 
In addition, when $w$ is considered, the tensions will be further reduced, as shown by the $\Omega_m$--$H_0$ and the 
$\Omega_m$--$\sigma_8$ panels. 
In order to quantify how the data consistencies are improved by the sterile neutrino and dynamical dark energy, we give the fit results 
of $H_0$, $\sigma_8(\Omega_m/0.27)^{0.3}$, $\sigma_8(\Omega_m/0.27)^{0.46}$, and $r_{0.002}$ for the $\Lambda$CDM+$r$, the 
$\Lambda$CDM+$r$+$N_{\rm eff}$+$m_{\nu,{\rm sterile}}^{\rm eff}$, the $w$CDM+$r$, and the $w$CDM+$r$+$N_{\rm eff}$+$m_{\nu,{\rm sterile}}^{\rm eff}$ models in Table~\ref{tab01}. 
The levels of tensions between fit results and observational results are also given. 
Under the constraints from Planck+WP+BAO, considering sterile neutrino improves the tension with $H_0$ from 2.4$\sigma$ to 1.0$\sigma$, 
the tension with SZ cluster counts from 4.3$\sigma$ to 2.0$\sigma$, and the tension with cosmic shear from 2.3$\sigma$ to 1.7$\sigma$, respectively; 
considering dynamical dark energy (with constant $w$) improves the above levels of tensions to 1.6$\sigma$, 3.2$\sigma$, and 2.3$\sigma$, respectively;
simultaneous consideration of sterile neutrino and dynamical dark energy improves the tension levels to 0.6$\sigma$, 1.7$\sigma$, and 1.5$\sigma$, respectively.
But in this case dynamical dark energy nearly does not affect the fit result of $r_{0.002}$. 

From the above analysis, we find that considering dynamical dark energy with constant $w$ could further improve the data consistency. 
In the $w$CDM+$r$+$\sum m_\nu$ model, the CMB+BAO data are basically consistent with $H_0$, SZ cluster counts, and cosmic shear data, but still in tension with the BICEP2 data. 
In the $w$CDM+$r$+$N_{\rm eff}$ model, the CMB+BAO data are consistent with the $H_0$ measurement, but still in some tension with 
cluster data and BICEP2 data (though the previous tensions are further reduced to some extent). 
In the $w$CDM+$r$+$\sum m_\nu$+$N_{\rm eff}$ model and the $w$CDM+$r$+$N_{\rm eff}$+$m_{\nu,{\rm sterile}}^{\rm eff}$ model, the 
CMB+BAO data are consistent with all the data sets considered, and thus these data sets can be combined together to perform joint constraints on these two models 
almost without any question.

\subsection{Constraints on dark energy, dark radiation, and neutrinos}

In this subsection, we give the fit results for the four models. Of course, we focus on the constraint results of dark energy, dark radiation, and neutrinos.

\begin{figure}[tbp]
\centering 
\includegraphics[scale=0.5]{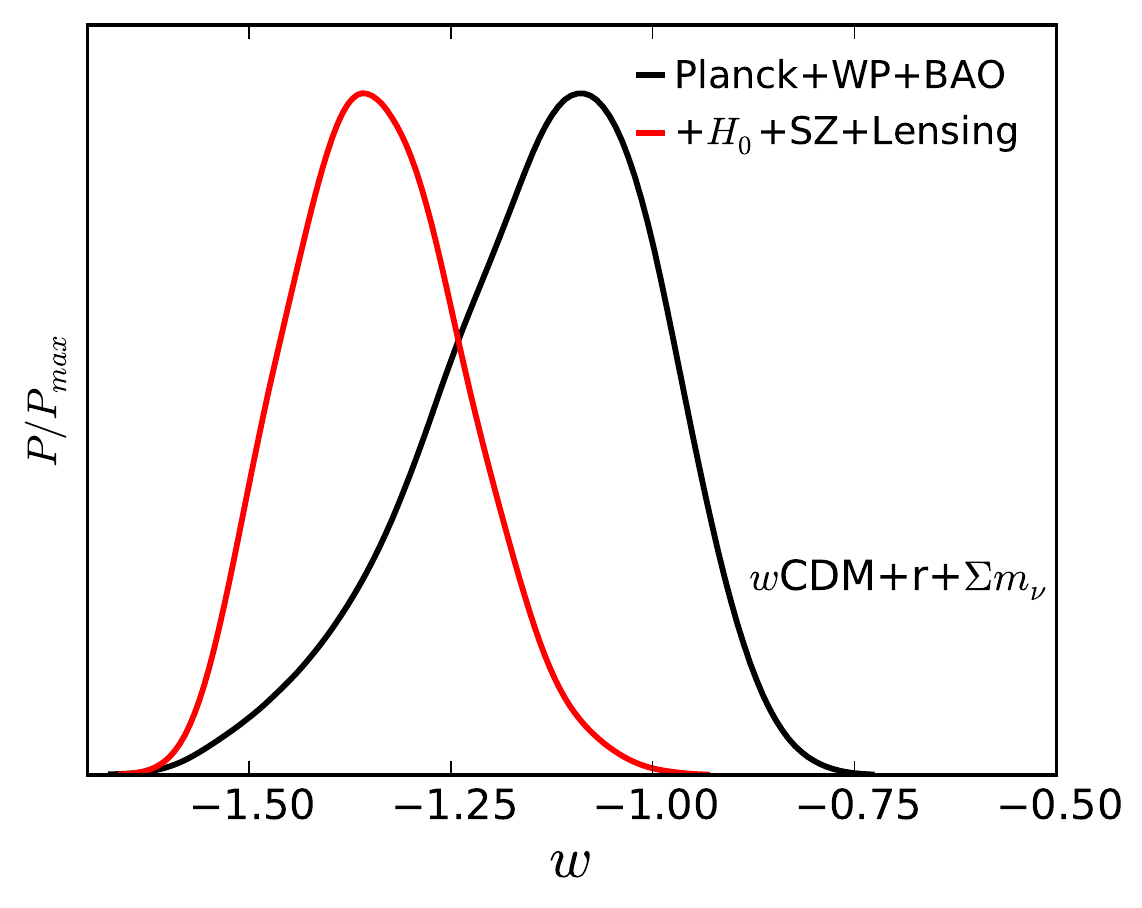}
\includegraphics[scale=0.5]{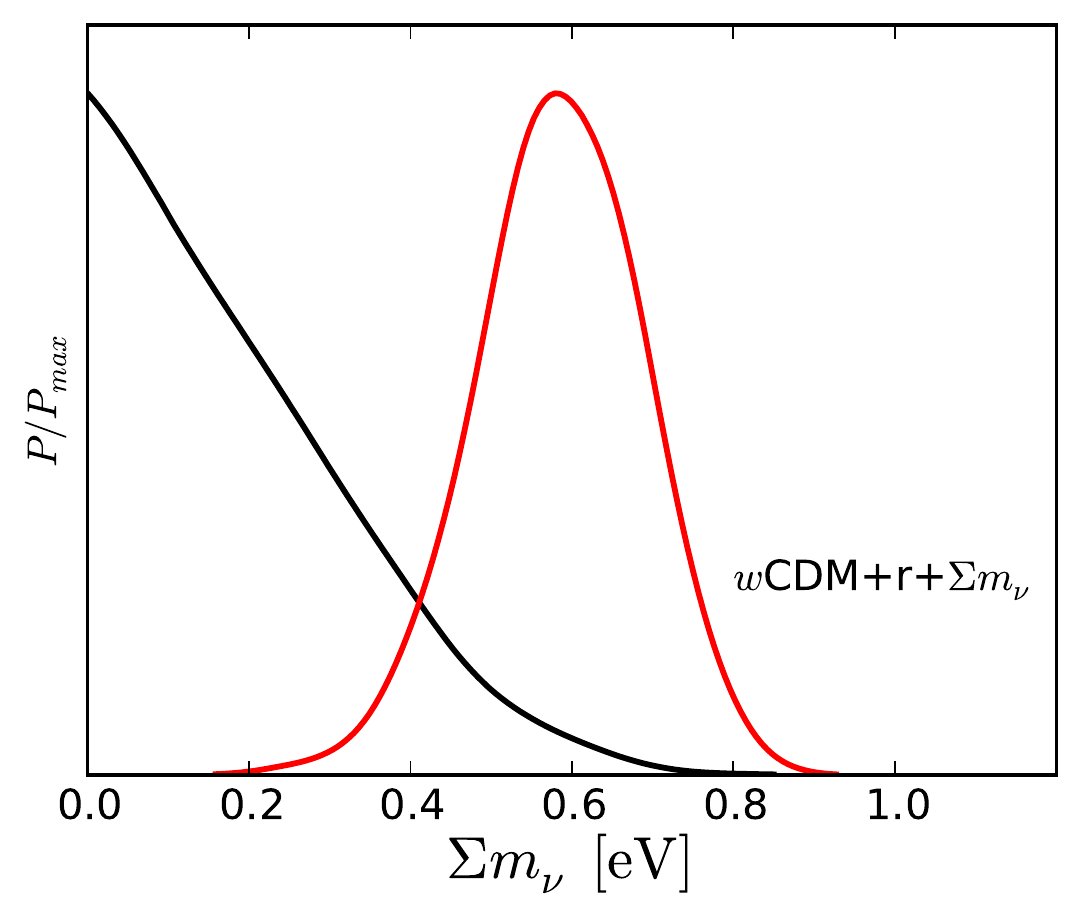}
\includegraphics[scale=0.5]{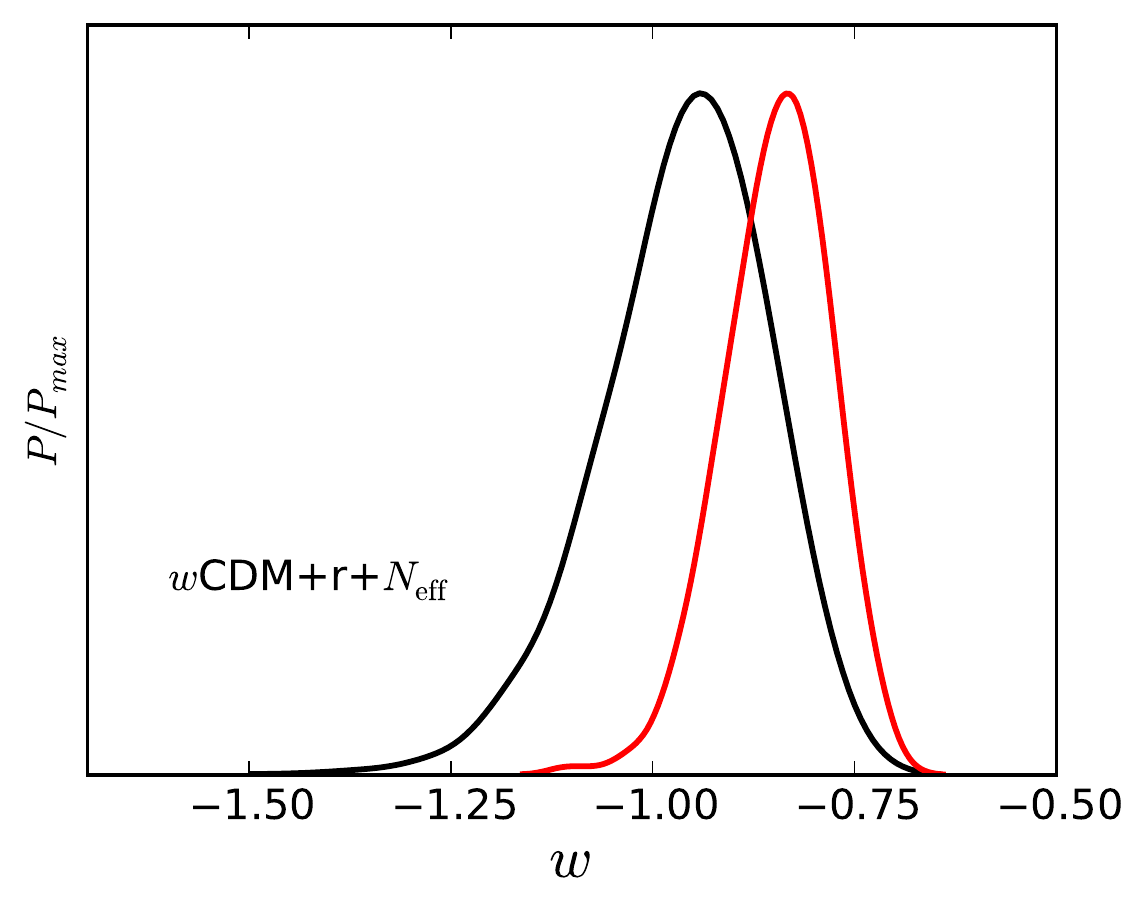}
\includegraphics[scale=0.5]{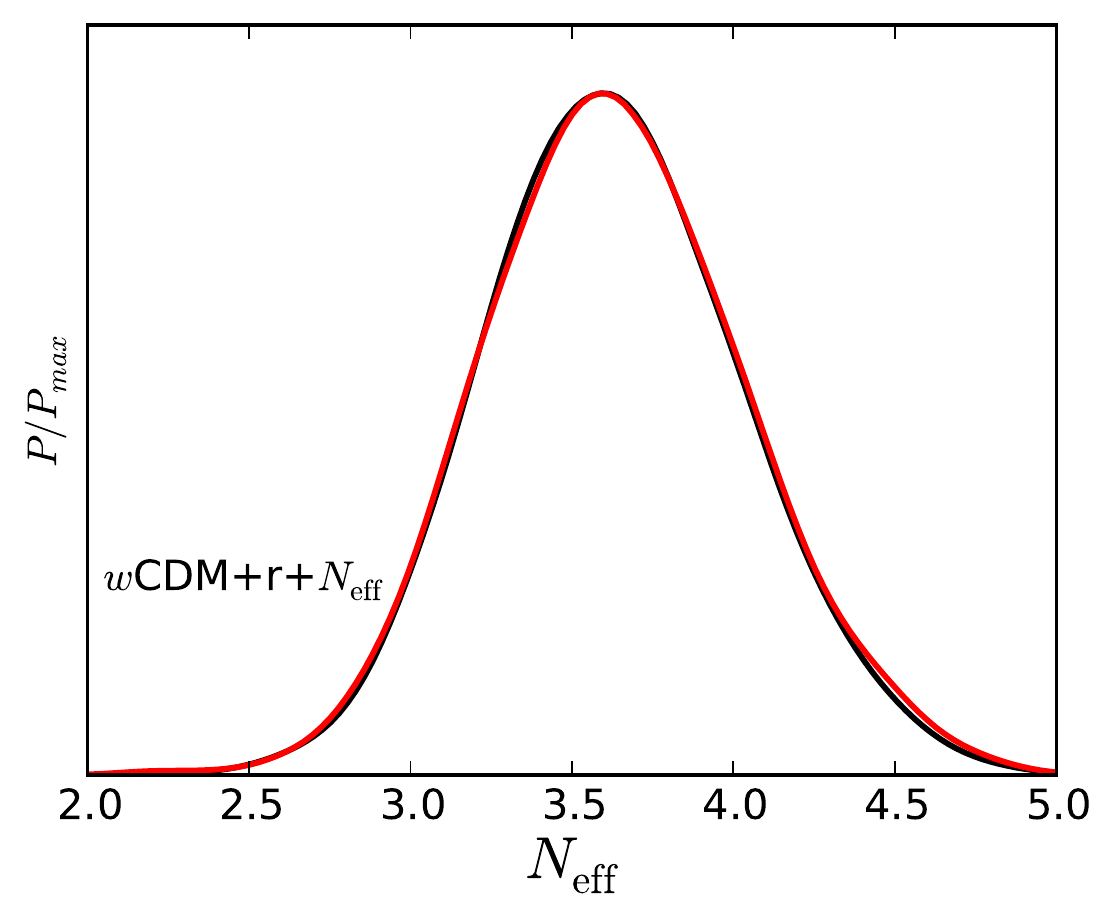}
\hfill
\caption{\label{fig12} One-dimensional posterior distributions for the $w$CDM+$r$+$\sum m_\nu$ model (upper) and the 
$w$CDM+$r$+$N_{\rm eff}$ model (lower).}
\end{figure}

Figure~\ref{fig12} shows the fit results for the $w$CDM+$r$+$\sum m_\nu$ model and the $w$CDM+$r$+$N_{\rm eff}$ model. 
In this figure, the one-dimensional posterior possibility distributions of $w$ and $\sum m_\nu$ are plotted for the $w$CDM+$r$+$\sum m_\nu$ model, 
and those of $w$ and $N_{\rm eff}$ are plotted for the $w$CDM+$r$+$N_{\rm eff}$ model. Black curves are from the Planck+WP+BAO constraints, and 
red curves are from the Planck+WP+BAO+$H_0$+SZ+Lensing constraints. 

For the $w$CDM+$r$+$\sum m_\nu$ model, the CMB+BAO data combination gives the constraint results: 
\begin{eqnarray}
w= -1.14^{+0.17}_{-0.11}, \nonumber \\
\sum m_\nu < 0.48~{\rm eV}.
\end{eqnarray}
Here, we show the $\pm 1\sigma$ errors for $w$, and show the 95\% CL upper limit for $\sum m_\nu$. 
Note that throughout the paper we quote $\pm 1\sigma$ errors, but when the parameters cannot be well constrained we only give the 2$\sigma$ upper limits for these parameters. 
The constraints on $\Omega_m$ and $H_0$ are: $\Omega_m=0.293^{+0.021}_{-0.018}$ and $H_0=70.1^{+2.3}_{-3.2}$ km s$^{-1}$ Mpc$^{-1}$. 
Since the CMB+BAO data are basically consistent with the other astrophysical data (note that in this subsection we do not discuss BICEP2), 
one can also combine these data together. For convenience, hereafter we also use ``other'' to denote the data combination of $H_0$+SZ+Lensing.  
The CMB+BAO+other data combination leads to:
\begin{eqnarray}
w=-1.34^{+0.10}_{-0.12}, \nonumber\\
\sum m_\nu=0.58^{+0.11}_{-0.10}~{\rm eV}.
\end{eqnarray}

One can clearly see that once the other data sets are combined, the value of $w$ becomes smaller. 
In fact, even though the consideration of dynamical dark energy could relieve the tension of $H_0$, the strong 
degeneracy between $\Omega_m$ and $H_0$ still exists; recall the discussion in the last subsection. 
So, when the other data sets are added, the $\Omega_m$--$H_0$ degeneracy is partly broken, resulting in that the 
$\Omega_m$ becomes smaller and $H_0$ becomes larger: $\Omega_m=0.272^{+0.013}_{-0.014}$ and 
$H_0=73.3\pm 2.0$ km s$^{-1}$ Mpc$^{-1}$. 
Since $w$ and $H_0$ are in anti-correlation, larger $H_0$ leads to smaller $w$. 
This is why $w$ becomes smaller in this case.
In addition, the neutrino mass is sensitive to the cluster data, and so the combination with SZ and Lensing data could tighten the constraint on $\sum m_\nu$.

For the $w$CDM+$r$+$N_{\rm eff}$ model, the CMB+BAO data combination gives:
\begin{eqnarray}
w= -0.96^{+0.12}_{-0.09}, \nonumber \\
N_{\rm eff}=3.63^{+0.38}_{-0.42}.
\end{eqnarray}
In this case, we also have $\Omega_m=0.306^{+0.017}_{-0.015}$ and $H_0=70.1^{+2.0}_{-2.4}$ km s$^{-1}$ Mpc$^{-1}$.
From the discussion in the last subsection, we learn that the CMB+BAO data are not well consistent with the SZ cluster counts data, and so in principle 
it is not appropriate to combine them together. 
If we combine these data together (CMB+BAO+other), we get the results:
\begin{eqnarray}
w= -0.85^{+0.08}_{-0.06}, \nonumber \\
N_{\rm eff}=3.63^{+0.39}_{-0.43}.
\end{eqnarray}
We must keep in mind that the above results are from the tensioned data combination and thus are not reliable. 
Since the SZ cluster counts data give a lower $\sigma_8$, we obtain a much higher $w$ in this case (recall that $w$ and $\sigma_8$ 
are in anti-correlation; see the relevant discussion in the last subsection). 
$N_{\rm eff}$ is insensitive to the cluster and cosmic shear data, and so in this case the constraint on $N_{\rm eff}$ is not changed.
In this case, we have $\Omega_m=0.306\pm 0.012$ and $H_0=68.7\pm 1.5$ km s$^{-1}$ Mpc$^{-1}$.

\begin{figure}[tbp]
\centering 
\includegraphics[scale=0.7]{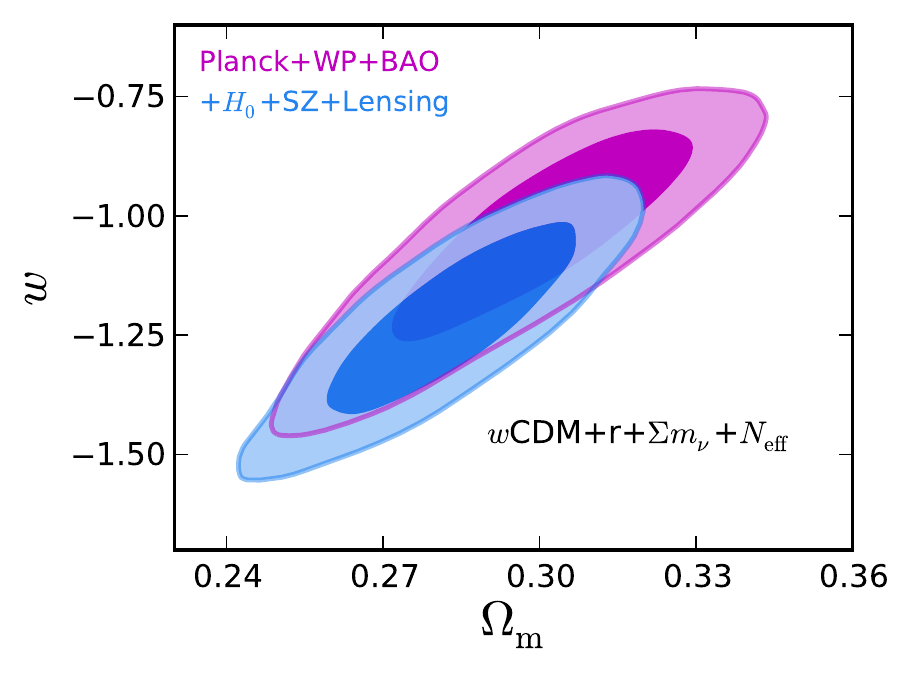}
\includegraphics[scale=0.7]{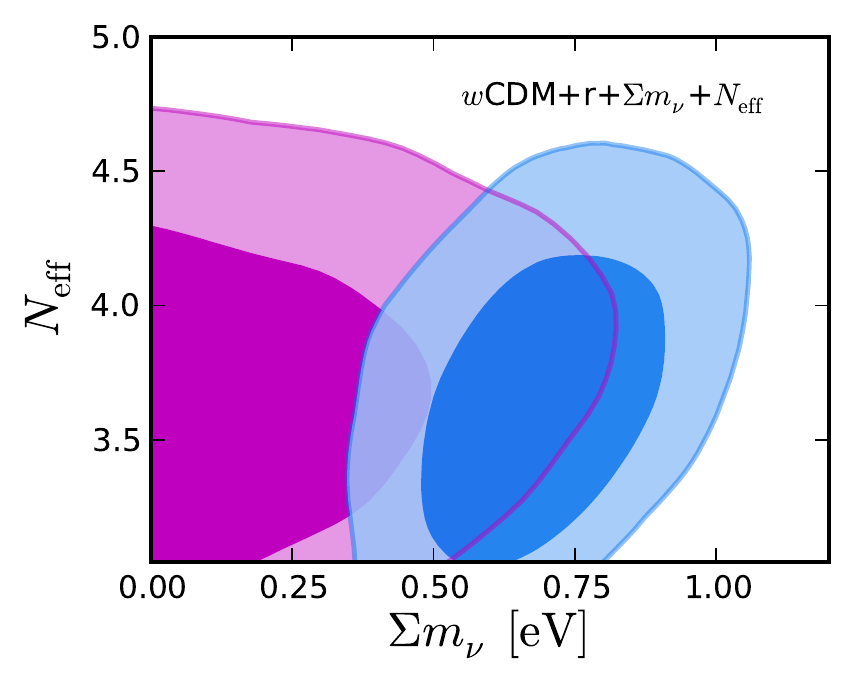}
\includegraphics[scale=0.7]{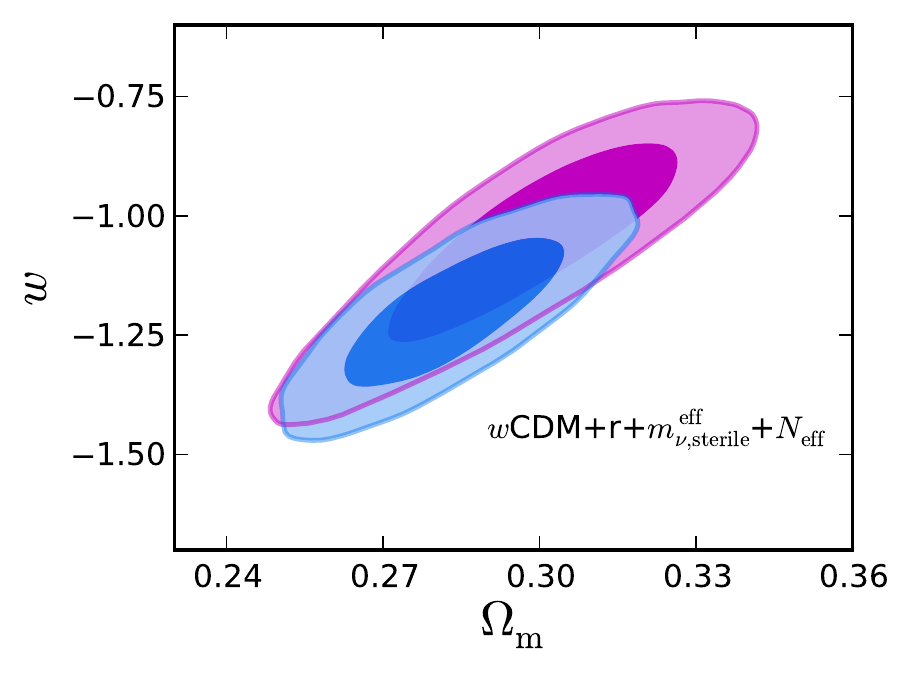}
\includegraphics[scale=0.7]{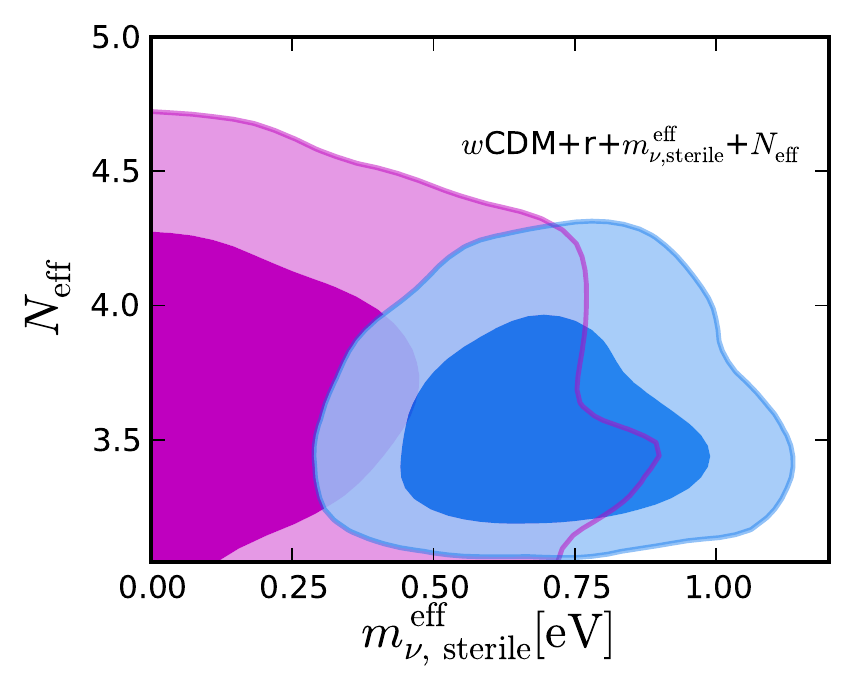}
\hfill
\caption{\label{fig13} Two-dimensional joint, marginalized constraints for the $w$CDM+$r$+$\sum m_\nu$+$N_{\rm eff}$ model (upper) 
and the $w$CDM+$r$+$N_{\rm eff}$+$m_{\nu,{\rm sterile}}^{\rm eff}$ model (lower).}
\end{figure}

Figure~\ref{fig13} summarizes the main results for the $w$CDM+$r$+$\sum m_\nu$+$N_{\rm eff}$ model and 
the $w$CDM+$r$+$N_{\rm eff}$+$m_{\nu,{\rm sterile}}^{\rm eff}$ model.
In this figure, the two-dimensional posterior possibility distribution contours in the $\Omega_m$--$w$ and $\sum m_\nu$--$N_{\rm eff}$ planes 
are plotted for the active neutrino model, and those in the $\Omega_m$--$w$ and $m_{\nu,{\rm sterile}}^{\rm eff}$--$N_{\rm eff}$ planes 
are plotted for the sterile neutrino model. 
Purple contours are from the CMB+BAO constraints, and blue contours are from the CMB+BAO+other constraints. 
As discussed in the last subsection, in these two models the CMB+BAO data are consistent with the other observations considered in this paper, 
and thus the combination of these data sets is viewed as appropriate. In the following we report the fit results for the two models. 

For the $w$CDM+$r$+$\sum m_\nu$+$N_{\rm eff}$ model, the CMB+BAO data give the fit results:
\begin{eqnarray}
w= -1.05^{+0.17}_{-0.11}, \nonumber \\
\sum m_\nu<0.651~{\rm eV},~~~
N_{\rm eff}=3.68^{+0.38}_{-0.45}. 
\end{eqnarray}
In this case, we have $\Omega_m=0.299^{+0.021}_{-0.016}$ and $H_0=71.6^{+2.4}_{-3.2}$ km s$^{-1}$ Mpc$^{-1}$. 
Furthermore, the CMB+BAO+other data give the fit results:
\begin{eqnarray}
w= -1.22^{+0.14}_{-0.12}, \nonumber \\
\sum m_\nu=0.69^{+0.13}_{-0.15}~{\rm eV},~~~
N_{\rm eff}=3.66^{+0.35}_{-0.41}. 
\end{eqnarray}
In this case, we have $\Omega_m=0.283\pm 0.015$ and $H_0=74.1\pm 2.0$ km s$^{-1}$ Mpc$^{-1}$.
From the upper panels of Fig.~\ref{fig13}, one can see that the CMB+BAO data are well consistent with the other astrophysical data for this model.

For the $w$CDM+$r$+$N_{\rm eff}$+$m_{\nu,{\rm sterile}}^{\rm eff}$ model, the CMB+BAO data combination gives the constraint results:
\begin{eqnarray}
w= -1.06^{+0.16}_{-0.11}, \nonumber \\
N_{\rm eff}=3.68^{+0.38}_{-0.45},~~~
m_{\nu,{\rm sterile}}^{\rm eff}<0.70~{\rm eV}.
\end{eqnarray}
In this case, we have $\Omega_m=0.298^{+0.020}_{-0.017}$ and $H_0=71.9^{+2.3}_{-3.2}$ km s$^{-1}$ Mpc$^{-1}$. 
Furthermore, the CMB+BAO+other data give the constraint results:
\begin{eqnarray}
w= -1.20\pm 0.10, \nonumber \\
N_{\rm eff}=3.56^{+0.15}_{-0.33},~~~
m_{\nu,{\rm sterile}}^{\rm eff}=0.70\pm 0.16~{\rm eV}.
\end{eqnarray}
In this case, we have $\Omega_m=0.283^{+0.013}_{-0.014}$ and $H_0=73.7\pm 2.0$ km s$^{-1}$ Mpc$^{-1}$. 
The same to the above active neutrino model, we also find from the lower panel of Fig.~\ref{fig13} that the CMB+BAO data are well 
 consistent with the other observations for the sterile neutrino model. 

We find that the $w$CDM-based neutrino models fit the cosmological data much better than the corresponding $\Lambda$CDM-based models. 
The $\chi_{\rm min}^2$ values of the $w$CDM-based models are given in Tables~\ref{tab1}--\ref{tab4} in Appendix \ref{aa}, 
and those of the corresponding $\Lambda$CDM-based models can be found in Tables 1--4 of Ref.~\cite{sterile2}. 
Note that in these tables actually the values of $-\ln {\cal L}_{\rm max}=\chi_{\rm min}^2/2$ are given. 
Fits to the CMB+BAO+other data lead to an increase of $\Delta\chi^2=-10.54$ for the $w$CDM+$r$+$\sum m_\nu$ model, 
$\Delta\chi^2=-3.64$ for the $w$CDM+$r$+$N_{\rm eff}$ model, $\Delta\chi^2=-4.28$ for the $w$CDM+$r$+$\sum m_\nu$+$N_{\rm eff}$ model, and 
$\Delta\chi^2=-7.5$ for the $w$CDM+$r$+$N_{\rm eff}$+$m_{\nu,{\rm sterile}}^{\rm eff}$ model, compared to the corresponding $\Lambda$CDM-based models. 
According to the Akaike information criterion, if $\chi_{\rm min}^2$ improves by 2 or more with one additional parameter, its incorporation is justified. 
In this context, the performance of $w$CDM-based models (with only one more additional parameter beyond the corresponding $\Lambda$CDM-based models) 
improves $\chi_{\rm min}^2$ significantly. 

Furthermore, to see how the additional neutrino/dark radiation parameters improve the fits in the framework of $w$CDM, 
we report the $\chi^2$ value of fitting the $w$CDM+$r$ model to the CMB+BAO+other data, $\chi_{\rm min}^2=9841.00$. 
Therefore, compared to this model, we get $\Delta\chi^2=-18.12$ for the $w$CDM+$r$+$\sum m_\nu$ model, 
$\Delta\chi^2=-1.36$ for the $w$CDM+$r$+$N_{\rm eff}$ model, $\Delta\chi^2=-18.42$ for the $w$CDM+$r$+$\sum m_\nu$+$N_{\rm eff}$ model, and 
$\Delta\chi^2=-22.02$ for the $w$CDM+$r$+$N_{\rm eff}$+$m_{\nu,{\rm sterile}}^{\rm eff}$ model. 
It is clear to see that the neutrino mass plays a more important role than $N_{\rm eff}$ in the fits.

\subsection{Early-universe parameters and BICEP2}

\begin{figure}[tbp]
\centering 
\includegraphics[scale=0.7]{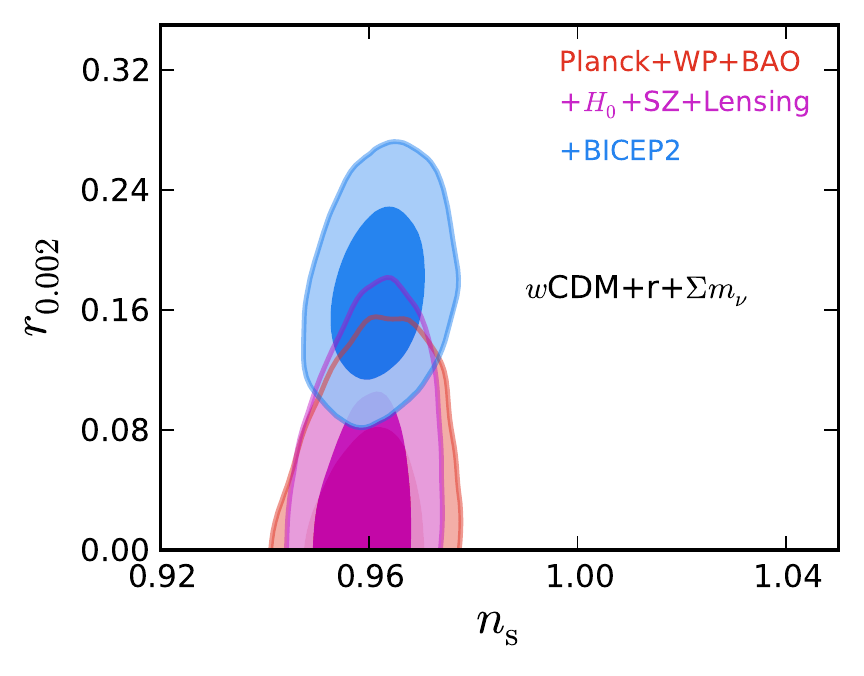}
\includegraphics[scale=0.7]{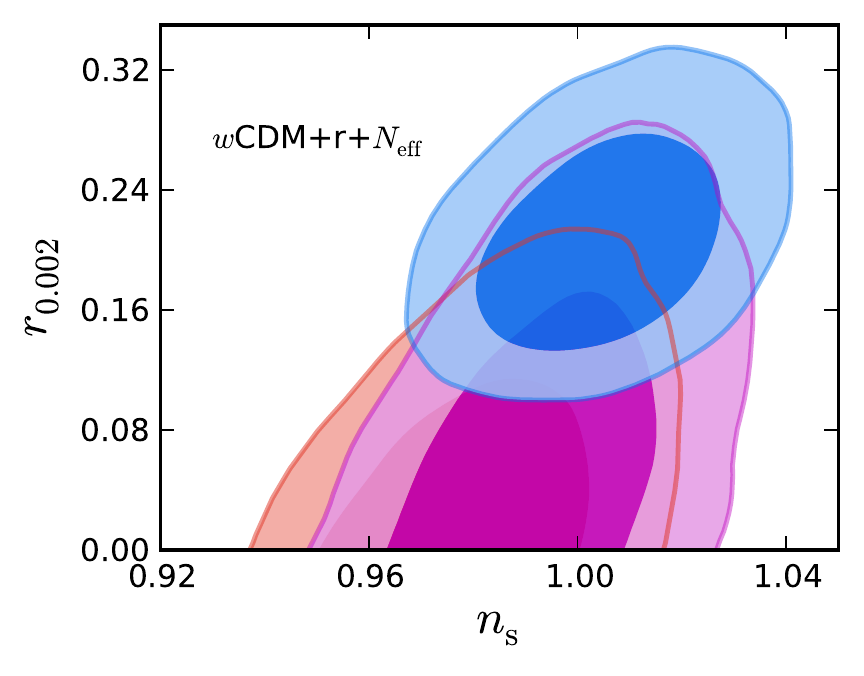}
\includegraphics[scale=0.7]{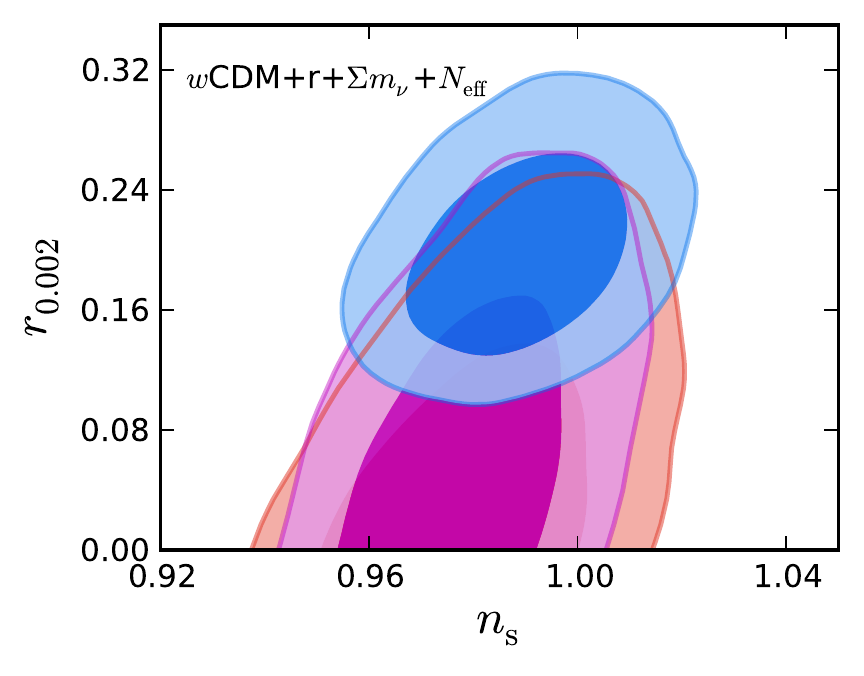}
\includegraphics[scale=0.7]{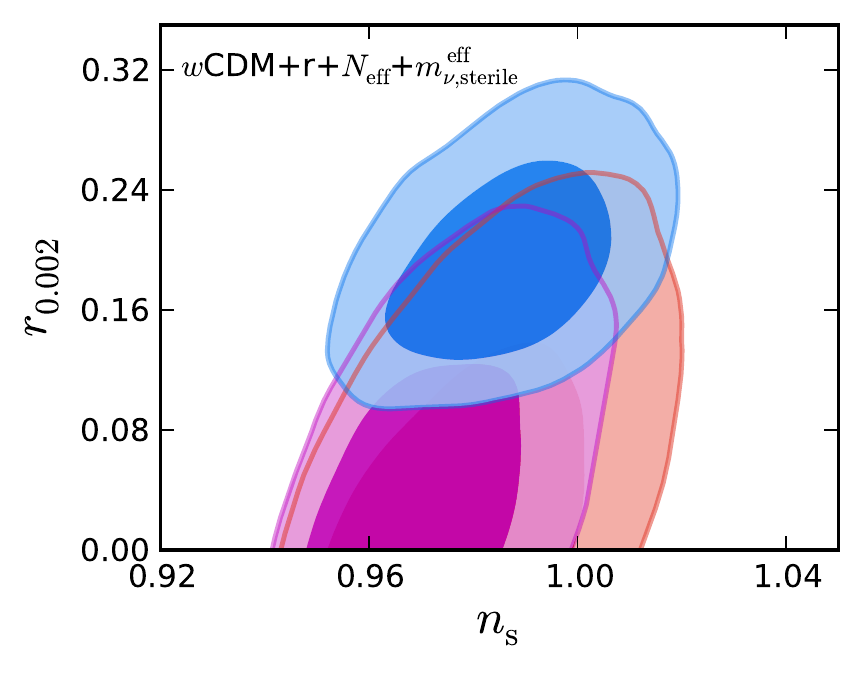}
\hfill
\caption{\label{fig14} Two-dimensional joint, marginalized constraints in the $r_{0.002}-n_s$ plane for the four models.}
\end{figure}

In this subsection we consider the constraints on $n_s$ and $r$ and the BICEP2 data. 
Figure~\ref{fig14} shows the two-dimensional joint constraints on the four model in the $n_s$--$r_{0.002}$ plane. 
The orange contours are for the CMB+BAO data, the purple ones are for the CMB+BAO+other data, and the 
blue ones are for the CMB+BAO+other+BICEP2 data.

For the $w$CDM+$r$+$\sum m_\nu$ model and the $w$CDM+$r$+$N_{\rm eff}$ model, since the tension between Planck 
and BICEP2 cannot be greatly reduced, we are not interested in these two models for this aspect. We thus only give the 95\% CL upper limits 
of $r_{0.002}$ given by the CMB+BAO data for these two models: $r_{0.002}<0.125$ for the $w$CDM+$r$+$\sum m_\nu$ model 
and $r_{0.002}<0.165$ for the $w$CDM+$r$+$N_{\rm eff}$ model. 

In the $w$CDM+$r$+$\sum m_\nu$+$N_{\rm eff}$ model and 
the $w$CDM+$r$+$N_{\rm eff}$+$m_{\nu,{\rm sterile}}^{\rm eff}$ model, the tension of $r$ between Planck and BICEP2 can be well 
relieved, and so we are interested in these two cases. 
For the $w$CDM+$r$+$\sum m_\nu$+$N_{\rm eff}$ model, the CMB+BAO data give:
\begin{equation}
n_s=0.983^{+0.015}_{-0.017},~~~r_{0.002}<0.200.
\end{equation}
Furthermore, the CMB+BAO+other+BICEP2 give the results:
\begin{equation}
n_s=0.989\pm 0.013,~~~
r_{0.002}=0.200^{+0.039}_{-0.049}.
\end{equation}
For the $w$CDM+$r$+$N_{\rm eff}$+$m_{\nu,{\rm sterile}}^{\rm eff}$ model, the CMB+BAO data give:
\begin{equation}
n_s=0.984^{+0.014}_{-0.017},~~~r_{0.002}<0.199.
\end{equation}
Furthermore, the CMB+BAO+other+BICEP2 give the results:
\begin{equation}
n_s=0.987^{+0.014}_{-0.013},~~~
r_{0.002}=0.193^{+0.038}_{-0.050}.
\end{equation}

At last, we discuss the goodness of fits to the full data combination (with BICEP2 involved) for these two models. 
To see the role the parameter $w$ plays in the fits, we take the $\Lambda$CDM-based models as references, and 
we obtain $\Delta\chi^2=-1.78$ for the $w$CDM+$r$+$\sum m_\nu$+$N_{\rm eff}$ model, and 
$\Delta\chi^2=-3.04$ for the $w$CDM+$r$+$N_{\rm eff}$+$m_{\nu,{\rm sterile}}^{\rm eff}$ model. 
Furthermore, to see how the two additional parameters relevant to neutrino/dark radiation improve the fits in the framework of $w$CDM, we 
choose the $w$CDM+$r$ model as a reference. In this case, we have $\chi_{\rm min}^2=9884.78$ for the $w$CDM+$r$ model. 
Therefore, we get $\Delta\chi^2=-20.44$ for the $w$CDM+$r$+$\sum m_\nu$+$N_{\rm eff}$ model, and 
$\Delta\chi^2=-20.18$ for the $w$CDM+$r$+$N_{\rm eff}$+$m_{\nu,{\rm sterile}}^{\rm eff}$ model.

\section{Conclusion}
\label{concl}

Based on the standard $\Lambda$CDM cosmology, the Planck data are in good agreement with the BAO data, but are in tension 
with other astrophysical observations, such as the $H_0$ direct measurement, the SZ cluster counts, the cosmic shear measurement, and so on \cite{planck}.
Some ingredients such as dark radiation, massive neutrinos, and dynamical dark energy were proposed to be considered in the cosmological model 
to reduce the tensions. In particular, extra sterile neutrino species, with additional parameters $N_{\rm eff}$ and $m_{\nu,{\rm sterile}}^{\rm eff}$, was 
used to resolve the tensions between Planck and other astrophysical observations. 
Then, after the detection of the B-mode polarization of the CMB by the BICEP2 experiment, which implies that the PGWs were possibly discovered, the base standard cosmology 
should at least be extended to the $\Lambda$CDM+$r$ model \cite{bicep2}. However, it was found that the BICEP2's result of $r$ is in tension with the Planck's fit result.
Again, it was proposed that the sterile neutrino can be used to reconcile the PGW results from BICEP2 and Planck \cite{zx14,WHu14}.
Other cases concerning neutrinos and dark radiation were also investigated in detail \cite{sterile2}.

In this paper, we study the neutrino cosmological models in which the cosmological constant is replaced with the dynamical dark energy with constant $w$. 
Four cases are considered, i.e., the $w$CDM+$r$+$\sum m_\nu$ model, the $w$CDM+$r$+$N_{\rm eff}$ model, the  $w$CDM+$r$+$\sum m_\nu$+$N_{\rm eff}$ model, and 
the $w$CDM+$r$+$N_{\rm eff}$+$m_{\nu,{\rm sterile}}^{\rm eff}$ model.
The observational data we consider in this paper include the Planck+WP, BAO, $H_0$, Planck SZ cluster, Planck CMB lensing, cosmic shear, and BICEP2 data. 
We first tested the consistency of these data sets in the four cosmological models, and then performed joint constraints on the properties of dark energy and neutrinos. 

For the $w$CDM+$r$+$\sum m_\nu$ model, the consideration of dynamical dark energy is rather helpful in relieving the tension of $H_0$, 
but even in this model $\Omega_m$ and $H_0$ are still in strong degeneracy; the dark energy parameter $w$ does not help much to reduce the tension between 
Planck and SZ cluster counts, and so only $\sum m_\nu$ plays a significant role in this aspect; also, $w$ could not affect the value of $r$.
Thus, the CMB+BAO data are basically consistent with other observations (except for BICEP2). The CMB+BAO constraint gives: 
$w=-1.14^{+0.17}_{-0.11}$ and $\sum m_\nu<0.48$ eV. The CMB+BAO+$H_0$+SZ+Lensing constraint gives: 
$w=-1.34^{+0.10}_{-0.12}$ and $\sum m_\nu=0.58^{+0.11}_{-0.10}$ eV.

For the $w$CDM+$r$+$N_{\rm eff}$ model, the dark energy parameter $w$ does not lead to more impact on $H_0$, but plays a significant role in 
reducing the tension of SZ cluster counts; larger $w$ leads to smaller $\sigma_8$ and larger $\Omega_m$, so some offset happens for the quantity 
$\sigma_8(\Omega_m/0.27)^{0.3}$, resulting in that the consideration of $w$ still could not well relieve the tension between Planck and clusters; 
$w$ can enhance the value of $r$ in this case (it is positively correlated with $n_s$), but the tension of $r$ still cannot be well relieved. 
The CMB+BAO data combination gives the results: $w=-0.96^{+0.12}_{-0.09}$, $N_{\rm eff}=3.63^{+0.38}_{-0.42}$, and $r_{0.002}<0.165$. 

The results of the active neutrinos plus dark radiation model  and the sterile neutrino model are very similar. 
So we only took the sterile neutrino model as an example to analyze the data consistency. 
We found that in this model all the tensions considered are greatly reduced. 
Under the constraint from CMB+BAO, for the $\Lambda$CDM+$r$ model, the tensions with $H_0$, SZ clusters, and cosmic shear are 
2.4$\sigma$, 4.3$\sigma$, and 2.3$\sigma$, respectively; 
for the $w$CDM+$r$+$N_{\rm eff}$+$m_{\nu,{\rm sterile}}^{\rm eff}$ model, the above tensions are reduced to 0.6$\sigma$, 1.7$\sigma$, and 1.5$\sigma$. 
The sterile neutrinos could well reconcile the $r$ results from BICEP2 and Planck, and in this case dynamical dark energy does not further improve the $r$ result. 

For the $w$CDM+$r$+$\sum m_\nu$+$N_{\rm eff}$ model, the CMB+BAO+$H_0$+SZ+Lensing constraint gives: 
$w= -1.22^{+0.14}_{-0.12}$, $\sum m_\nu=0.69^{+0.13}_{-0.15}~{\rm eV}$, and $N_{\rm eff}=3.66^{+0.35}_{-0.41}$; 
further adding the BICEP2 data, we have $n_s=0.989\pm 0.013$ and $r_{0.002}=0.200^{+0.039}_{-0.049}$.
For the $w$CDM+$r$+$N_{\rm eff}$+$m_{\nu,{\rm sterile}}^{\rm eff}$ model, the CMB+BAO+$H_0$+SZ+Lensing data 
combination leads to: $w= -1.20\pm 0.10$, $N_{\rm eff}=3.56^{+0.15}_{-0.33}$, and $m_{\nu,{\rm sterile}}^{\rm eff}=0.70\pm 0.16~{\rm eV}$.
Further adding the BICEP2 data gives: $n_s=0.987^{+0.014}_{-0.013}$ and $r_{0.002}=0.193^{+0.038}_{-0.050}$.

From our analysis, we found that both the $w$CDM+$r$+$\sum m_\nu$+$N_{\rm eff}$ model and the 
$w$CDM+$r$+$N_{\rm eff}$+$m_{\nu,{\rm sterile}}^{\rm eff}$ model are successful in relieving the tensions between Planck and 
other astrophysical observations. However, in the $w$CDM+$r$+$\sum m_\nu$+$N_{\rm eff}$ model, one has to assume that the 
massive active neutrinos coexist with dark radiation, and thus we believe that the sterile neutrino model is more natural. 
To further tightly constrain the property of dark energy, accurate type Ia supernova data are needed. 
We leave the further analysis on the sterile neutrinos plus dark energy model in future work.

\acknowledgments

We acknowledge the use of {\tt CosmoMC}. We thank Yun-He Li for helpful discussion.
JFZ is supported by the Provincial Department of Education of
Liaoning under Grant No. L2012087.
XZ is supported by the National Natural Science Foundation of
China under Grant No. 11175042 and the Fundamental Research Funds for the 
Central Universities under Grant No. N120505003.


\appendix
 \begin{appendices}


\section{Detailed constraint results}
\label{aa}

\begin{figure}[tbp]
\centering 
\includegraphics[scale=0.3]{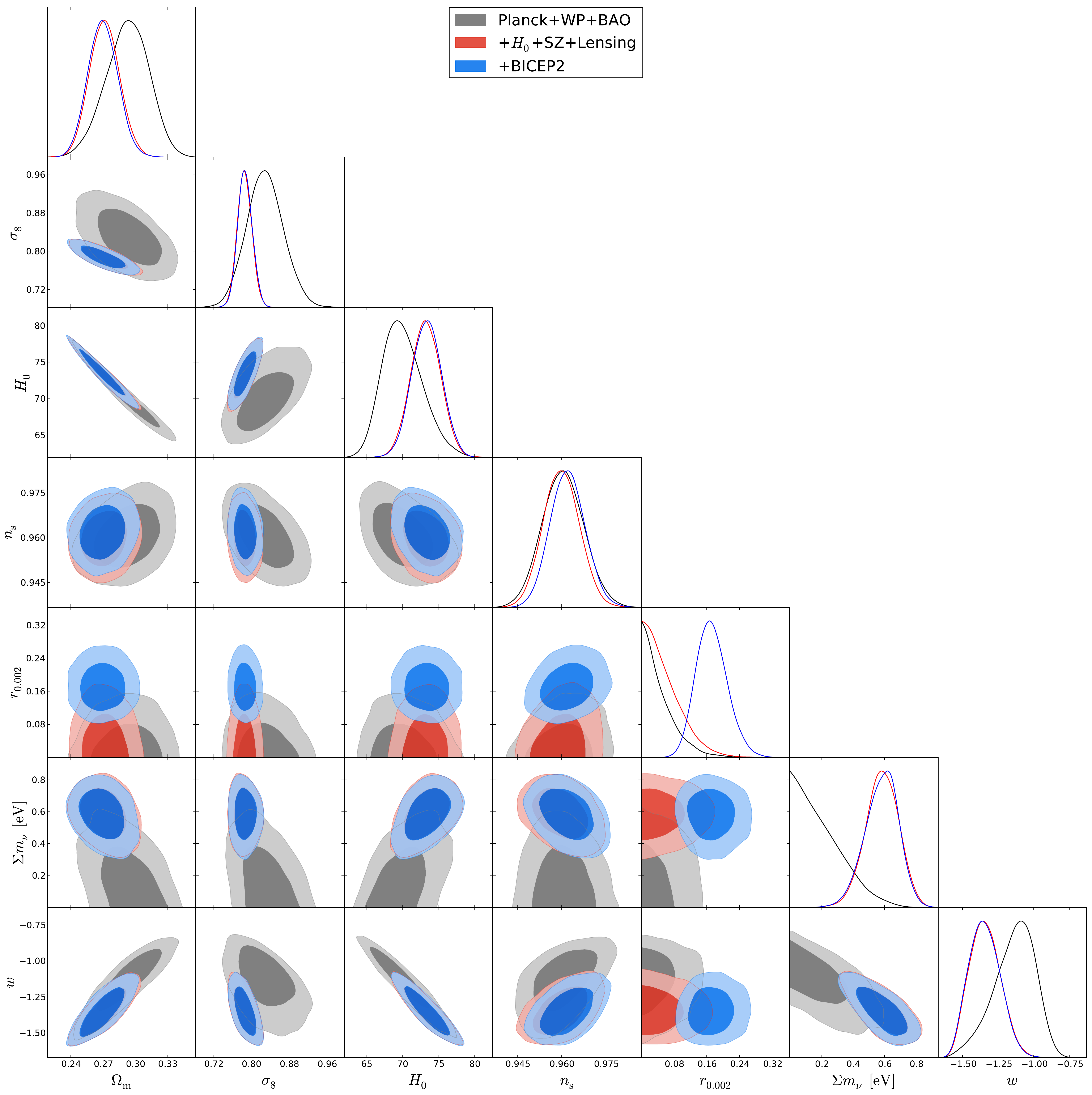}
\hfill
\caption{\label{fig1} Cosmological constraints on the $w$CDM+$r$+$\sum m_\nu$ model.}
\end{figure}

\begin{figure}[tbp]
\centering 
\includegraphics[scale=0.3]{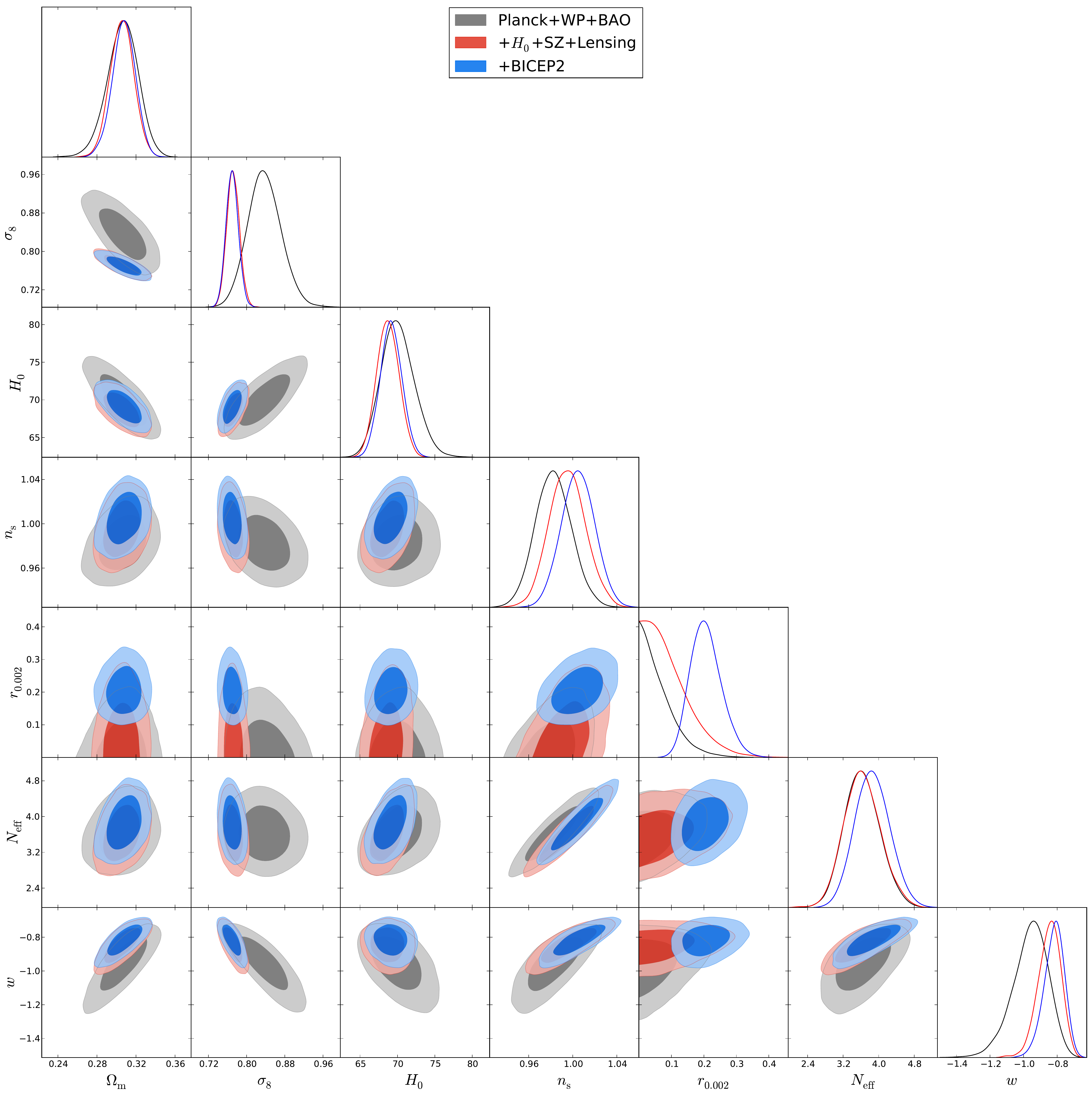}
\hfill
\caption{\label{fig2} Cosmological constraints on the $w$CDM+$r$+$N_{\rm eff}$ model.}
\end{figure}

\begin{figure}[tbp]
\centering 
\includegraphics[scale=0.26]{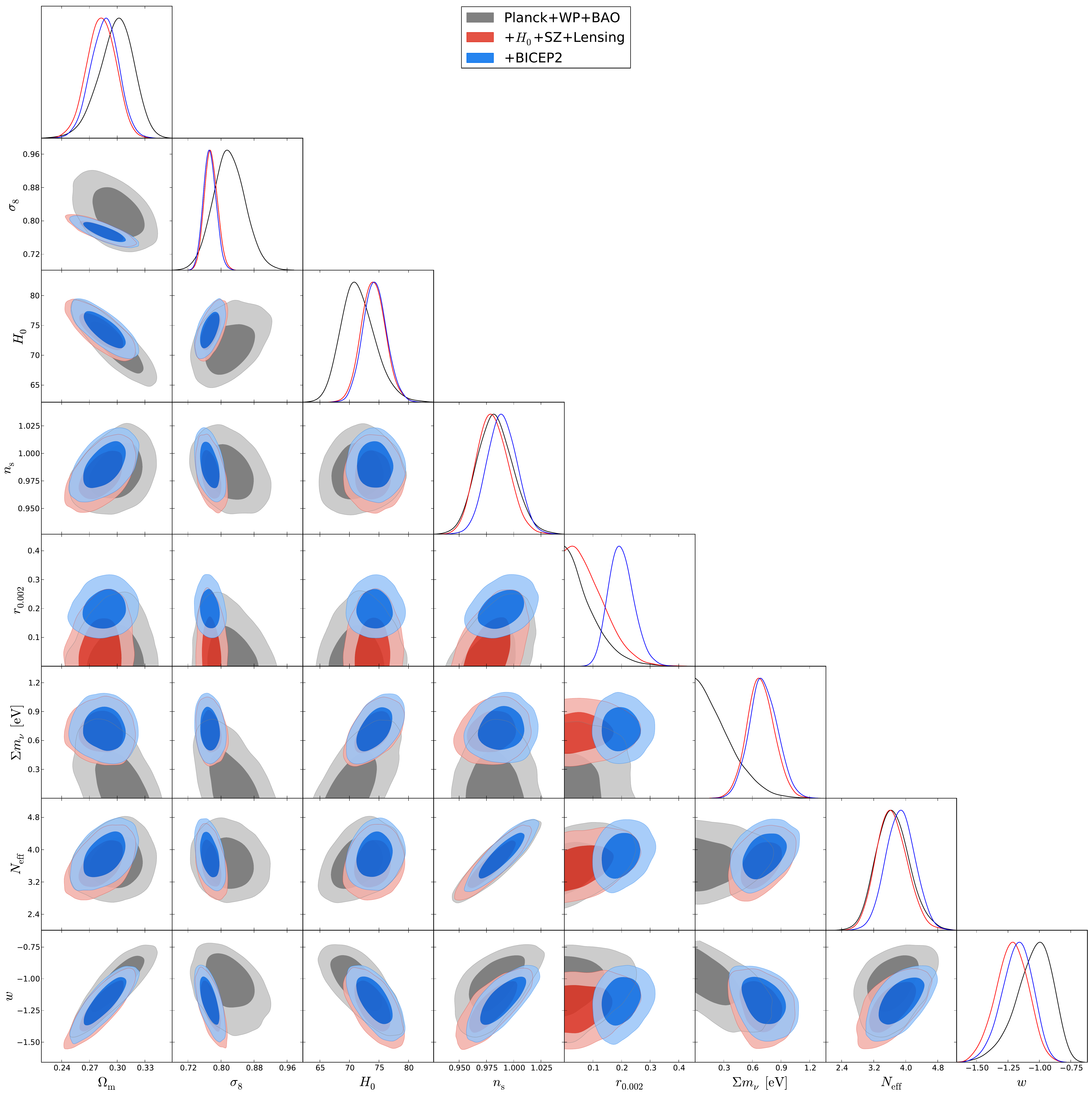}
\hfill
\caption{\label{fig3} Cosmological constraints on the $w$CDM+$r$+$\sum m_\nu$+$N_{\rm eff}$ model.}
\end{figure}

\begin{figure}[tbp]
\centering 
\includegraphics[scale=0.26]{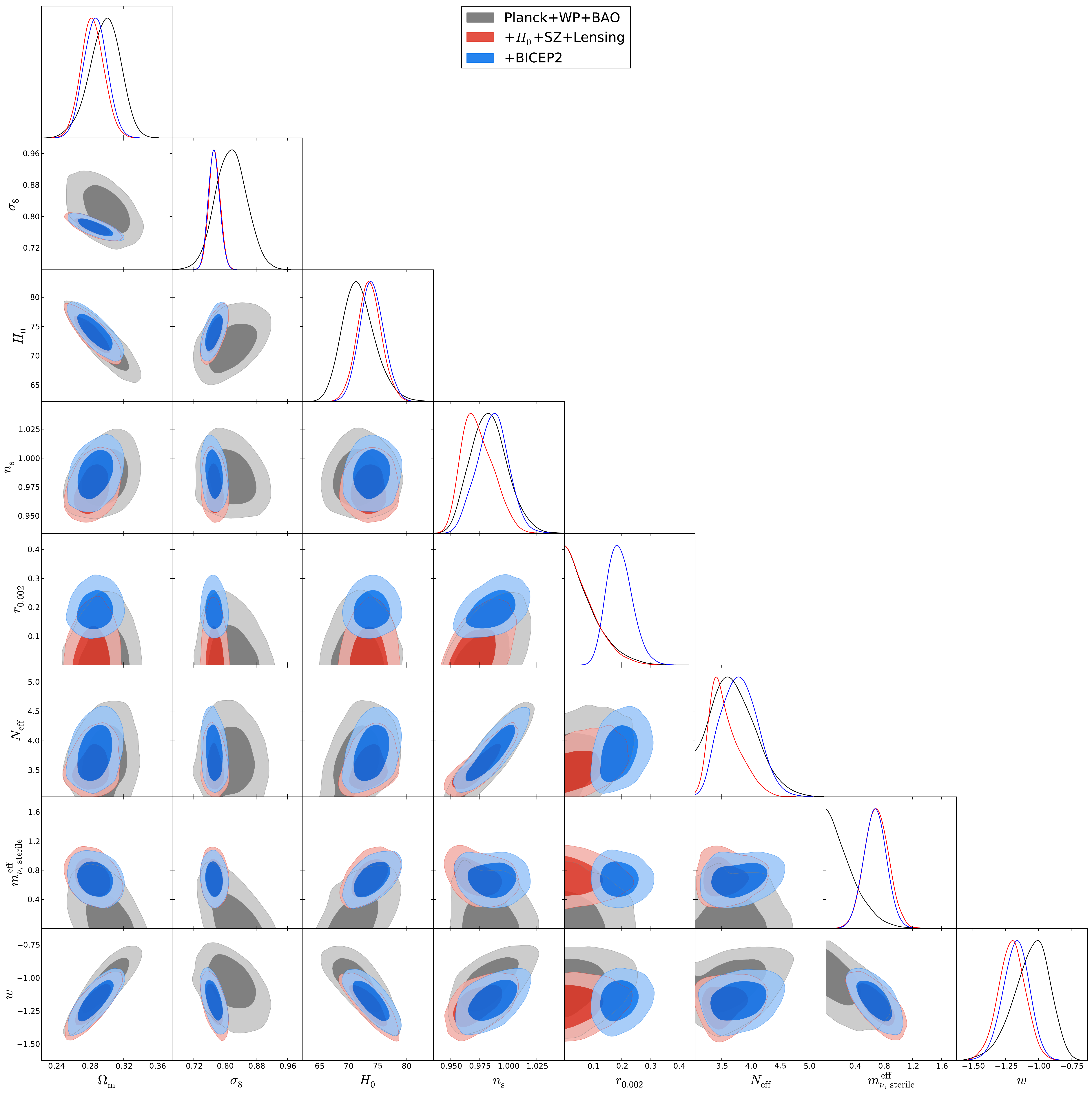}
\hfill
\caption{\label{fig4} Cosmological constraints on the $w$CDM+$r$+$N_{\rm eff}$+$m_{\nu,{\rm sterile}}^{\rm eff}$ model.}
\end{figure}

In this appendix, we give the detailed constraint results for the four cosmological models considered 
in this paper. The one- and two-dimensional joint, marginalized posterior probability distributions of the parameters 
for the models are shown in Figs.~\ref{fig1}--\ref{fig4}. 
Detailed fit values for the cosmological parameters are given in Tables~\ref{tab1}--\ref{tab4}.
In the tables, the $\pm 1\sigma$ errors are quoted, but for the parameters that cannot be well constrained, only the 
2$\sigma$ upper limits are given.

Figure~\ref{fig1} and Table~\ref{tab1} summarize the fit results for the $w$CDM+$r$+$\sum m_\nu$ model. 
Figure~\ref{fig2} and Table~\ref{tab2} summarize the fit results for the $w$CDM+$r$+$N_{\rm eff}$ model. 
Figure~\ref{fig3} and Table~\ref{tab3} summarize the fit results for the $w$CDM+$r$+$\sum m_\nu$+$N_{\rm eff}$ model. 
Figure~\ref{fig4} and Table~\ref{tab4} summarize the fit results for the $w$CDM+$r$+$N_{\rm eff}$+$m_{\nu,{\rm sterile}}^{\rm eff}$ model.

%

\begin{table}[tbp]\tiny
\centering
\begin{tabular}{|lccccccccc|}
\hline
 &\multicolumn{2}{c}{Planck+WP+BAO} & & \multicolumn{2}{c}{+$H_0$+SZ+Lensing} & & \multicolumn{2}{c}{+BICEP2} & \\
\cline{2-3}\cline{5-6}\cline{8-9}
Parameter & Best fit & 68\% limits && Best fit & 68\% limits && Best fit & 68\% limits & \\
\hline
$\Omega_bh^2$&$0.02236$&$0.02202^{+0.00027}_{-0.00026}$&&$0.02215$&$0.02211\pm0.00024$&&$0.02200$&$0.02205\pm0.00023$&\\
$\Omega_ch^2$&$0.1180$&$0.1193\pm0.0023$&&$0.1172$&$0.1171^{+0.0013}_{-0.0012}$&&$0.1176$&$0.1169^{+0.0013}_{-0.0012}$&\\
$100\theta_{\rm MC}$&$1.04130$&$1.04124\pm0.00059$&&$1.04103$&$1.04136^{+0.00055}_{-0.00056}$&&$1.04119$&$1.04136\pm0.00055$&\\
$\tau$&$0.098$&$0.090^{+0.012}_{-0.014}$&&$0.093$&$0.093^{+0.012}_{-0.014}$&&$0.092$&$0.093^{+0.012}_{-0.013}$&\\
$\Sigma m_\nu$&$0.039$&$<0.480$&&$0.580$&$0.584^{+0.111}_{-0.097}$&&$0.545$&$0.581^{+0.115}_{-0.094}$&\\
$w$&$-1.01$&$-1.14^{+0.17}_{-0.11}$&&$-1.30$&$-1.34^{+0.10}_{-0.12}$&&$-1.35$&$-1.34^{+0.10}_{-0.12}$&\\
$n_s$&$0.9651$&$0.9605\pm0.0068$&&$0.9632$&$0.9599\pm0.0059$&&$0.9623$&$0.9621^{+0.0057}_{-0.0058}$&\\
${\rm{ln}}(10^{10} A_s)$&$3.100$&$3.086^{+0.024}_{-0.025}$&&$3.088$&$3.087^{+0.023}_{-0.025}$&&$3.084$&$3.086^{+0.022}_{-0.025}$&\\
$r_{0.05}$&$0.010$&$<0.133$&&$0.051$&$<0.152$&&$0.180$&$0.181^{+0.034}_{-0.039}$&\\
\hline
$\Omega_\Lambda$&$0.701$&$0.707^{+0.018}_{-0.021}$&&$0.720$&$0.728^{+0.014}_{-0.013}$&&$0.730$&$0.730\pm0.014$&\\
$\Omega_m$&$0.299$&$0.293^{+0.021}_{-0.018}$&&$0.280$&$0.272^{+0.013}_{-0.014}$&&$0.270$&$0.270\pm0.014$&\\
$\sigma_8$&$0.835$&$0.830\pm0.037$&&$0.780$&$0.786\pm0.015$&&$0.797$&$0.787^{+0.015}_{-0.016}$&\\
$H_0$&$68.6$&$70.1^{+2.3}_{-3.2}$&&$72.0$&$73.3\pm2.0$&&$73.5$&$73.4\pm2.0$&\\
$r_{0.002}$&$0.0094$&$<0.125$&&$0.046$&$<0.143$&&$0.171$&$0.173^{+0.034}_{-0.042}$&\\
\hline
$-\ln\mathcal{L}_{\rm{max}}$ &\multicolumn{2}{c}{4904.82} & & \multicolumn{2}{c}{4911.44} & & \multicolumn{2}{c}{4934.33} & \\
\hline
\end{tabular}
\caption{\label{tab1} Fitting results for the $w$CDM+$r$+$\sum m_\nu$ model. We quote $\pm 1\sigma$ errors, but
for the parameters that cannot be well constrained, we quote the 95\% CL upper limits.}
\end{table}



\begin{table}[tbp]\tiny
\centering
\begin{tabular}{|lccccccccc|}
\hline
 &\multicolumn{2}{c}{Planck+WP+BAO} & & \multicolumn{2}{c}{+$H_0$+SZ+Lensing} & & \multicolumn{2}{c}{+BICEP2} & \\
\cline{2-3}\cline{5-6}\cline{8-9}
Parameter & Best fit & 68\% limits && Best fit & 68\% limits && Best fit & 68\% limits & \\
\hline
$\Omega_bh^2$&$0.02230$&$0.02245^{+0.00037}_{-0.0004}$&&$0.02278$&$0.02291^{+0.00038}_{-0.00037}$&&$0.02287$&$0.02302\pm0.00036$&\\
$\Omega_ch^2$&$0.1238$&$0.127^{+0.0054}_{-0.0061}$&&$0.1179$&$0.1207^{+0.0049}_{-0.0053}$&&$0.1214$&$0.1234\pm0.0048$&\\
$100\theta_{\rm MC}$&$1.04106$&$1.04057^{+0.00073}_{-0.00072}$&&$1.04095$&$1.04114^{+0.00071}_{-0.00072}$&&$1.04112$&$1.04092^{+0.00068}_{-0.00069}$&\\
$\tau$&$0.098$&$0.095^{+0.013}_{-0.016}$&&$0.087$&$0.091^{+0.014}_{-0.016}$&&$0.081$&$0.093^{+0.013}_{-0.015}$&\\
$w$&$-0.974$&$-0.962^{+0.120}_{-0.087}$&&$-0.886$&$-0.847^{+0.075}_{-0.059}$&&$-0.832$&$-0.819^{+0.067}_{-0.053}$&\\
$N_{\rm eff}$&$3.4$&$3.63^{+0.38}_{-0.42}$&&$3.41$&$3.63^{+0.39}_{-0.43}$&&$3.66$&$3.86^{+0.38}_{-0.41}$&\\
$n_s$&$0.975$&$0.983^{+0.016}_{-0.017}$&&$0.985$&$0.995\pm0.016$&&$0.996$&$1.005\pm0.015$&\\
${\rm{ln}}(10^{10}A_s)$&$3.116$&$3.115\pm0.032$&&$3.082$&$3.094^{+0.032}_{-0.035}$&&$3.078$&$3.104^{+0.03}_{-0.034}$&\\
$r_{0.05}$&$0.002$&$<0.158$&&$0.024$&$<0.203$&&$0.192$&$0.188^{+0.034}_{-0.042}$&\\
\hline
$\Omega_\Lambda$&$0.694$&$0.694^{+0.015}_{-0.017}$&&$0.701$&$0.694\pm0.012$&&$0.69$&$0.692\pm0.011$&\\
$\Omega_m$&$0.306$&$0.306^{+0.017}_{-0.015}$&&$0.299$&$0.306\pm0.012$&&$0.31$&$0.308\pm0.011$&\\
$\sigma_8$&$0.838$&$0.838^{+0.033}_{-0.037}$&&$0.776$&$0.772\pm0.013$&&$0.763$&$0.77\pm0.012$&\\
$H_0$&$69.3$&$70.1^{+2.0}_{-2.4}$&&$68.7$&$68.7\pm1.5$&&$68.4$&$69.2\pm1.4$&\\
$r_{0.002}$&$0.002$&$<0.165$&&$0.023$&$<0.224$&&$0.205$&$0.207^{+0.040}_{-0.054}$&\\
\hline
$-\ln\mathcal{L}_{\rm{max}}$ &\multicolumn{2}{c}{4903.64} & & \multicolumn{2}{c}{4919.82} & & \multicolumn{2}{c}{4940.82} & \\
\hline
\end{tabular}
\caption{\label{tab2} Fitting results for the $w$CDM+$r$+$N_{\rm eff}$ model. We quote $\pm 1\sigma$ errors, but
for the parameters that cannot be well constrained, we quote the 95\% CL upper limits.}
\end{table}


\begin{table}[tbp]\tiny
\centering
\begin{tabular}{|lccccccccc|}
\hline
 &\multicolumn{2}{c}{Planck+WP+BAO} & & \multicolumn{2}{c}{+$H_0$+SZ+Lensing} & & \multicolumn{2}{c}{+BICEP2} & \\
\cline{2-3}\cline{5-6}\cline{8-9}
Parameter & Best fit & 68\% limits && Best fit & 68\% limits && Best fit & 68\% limits & \\
\hline
$\Omega_bh^2$&$0.02217$&$0.02245^{+0.00038}_{-0.00041}$&&$0.02244$&$0.02253^{+0.00034}_{-0.00037}$&&$0.02272$&$0.02261^{+0.00034}_{-0.00035}$&\\
$\Omega_ch^2$&$0.1253$&$0.1274^{+0.0054}_{-0.0062}$&&$0.1224$&$0.1250^{+0.0047}_{-0.0055}$&&$0.1256$&$0.1276^{+0.0048}_{-0.0047}$&\\
$100\theta_{\rm MC}$&$1.04043$&$1.04047^{+0.00076}_{-0.00075}$&&$1.04095$&$1.04060^{+0.00071}_{-0.00072}$&&$1.04089$&$1.04036\pm0.00067$&\\
$\tau$&$0.096$&$0.095^{+0.014}_{-0.016}$&&$0.093$&$0.099^{+0.013}_{-0.016}$&&$0.104$&$0.102^{+0.013}_{-0.015}$&\\
$\Sigma m_\nu$&$0.052$&$<0.651$&&$0.61$&$0.69^{+0.13}_{-0.15}$&&$0.60$&$0.72^{+0.15}_{-0.14}$&\\
$w$&$-0.99$&$-1.05^{+0.17}_{-0.11}$&&$-1.19$&$-1.22^{+0.14}_{-0.12}$&&$-1.11$&$-1.18^{+0.14}_{-0.11}$&\\
$N_{\rm eff}$&$3.51$&$3.68^{+0.38}_{-0.45}$&&$3.52$&$3.66^{+0.35}_{-0.41}$&&$3.73$&$3.87\pm0.36$&\\
$n_s$&$0.979$&$0.983^{+0.015}_{-0.017}$&&$0.978$&$0.981^{+0.014}_{-0.015}$&&$0.988$&$0.989\pm0.013$&\\
${\rm{ln}}(10^{10}A_s)$&$3.115$&$3.116^{+0.031}_{-0.036}$&&$3.102$&$3.119^{+0.031}_{-0.035}$&&$3.131$&$3.129^{+0.03}_{-0.032}$&\\
$r_{0.05}$&$0.009$&$<0.188$&&$0.047$&$<0.202$&&$0.172$&$0.190^{+0.035}_{-0.04}$&\\
\hline
$\Omega_\Lambda$&$0.700$&$0.701^{+0.016}_{-0.021}$&&$0.720$&$0.717\pm0.015$&&$0.709$&$0.714^{+0.015}_{-0.016}$&\\
$\Omega_m$&$0.300$&$0.299^{+0.021}_{-0.016}$&&$0.280$&$0.283\pm0.015$&&$0.291$&$0.286^{+0.016}_{-0.015}$&\\
$\sigma_8$&$0.847$&$0.820\pm0.037$&&$0.771$&$0.776^{+0.015}_{-0.017}$&&$0.774$&$0.773^{+0.015}_{-0.016}$&\\
$H_0$&$70.3$&$71.6^{+2.4}_{-3.2}$&&$73.5$&$74.1\pm2.0$&&$72.9$&$74.3^{+1.9}_{-2.1}$&\\
$r_{0.002}$&$0.009$&$<0.200$&&$0.044$&$<0.213$&&$0.177$&$0.200^{+0.039}_{-0.049}$&\\
\hline
$-\ln\mathcal{L}_{\rm{max}}$ &\multicolumn{2}{c}{4904.45} & & \multicolumn{2}{c}{4911.29} & & \multicolumn{2}{c}{4932.17} & \\
\hline
\end{tabular}
\caption{\label{tab3} Fitting results for the $w$CDM+$r$+$\sum m_\nu$+$N_{\rm eff}$ model. We quote $\pm 1\sigma$ errors, but
for the parameters that cannot be well constrained, we quote the 95\% CL upper limits.}
\end{table}



\begin{table}[tbp]\tiny
\centering
\begin{tabular}{|lccccccccc|}
\hline
 &\multicolumn{2}{c}{Planck+WP+BAO} & & \multicolumn{2}{c}{+$H_0$+SZ+Lensing} & & \multicolumn{2}{c}{+BICEP2} & \\
\cline{2-3}\cline{5-6}\cline{8-9}
Parameter & Best fit & 68\% limits && Best fit & 68\% limits && Best fit & 68\% limits & \\
\hline
$\Omega_bh^2$&$0.02242$&$0.02248^{+0.00033}_{-0.00038}$&&$0.02232$&$0.02253^{+0.00028}_{-0.0003}$&&$0.02267$&$0.02261\pm0.00029$&\\
$\Omega_ch^2$&$0.1242$&$0.1275^{+0.0056}_{-0.0061}$&&$0.1192$&$0.1228^{+0.0035}_{-0.005}$&&$0.1267$&$0.1265^{+0.0047}_{-0.0048}$&\\
$100\theta_{\rm MC}$&$1.04065$&$1.04045^{+0.00075}_{-0.00074}$&&$1.04134$&$1.04075^{+0.00066}_{-0.00065}$&&$1.04035$&$1.04044\pm0.00068$&\\
$\tau$&$0.095$&$0.096^{+0.013}_{-0.016}$&&$0.087$&$0.095^{+0.013}_{-0.016}$&&$0.099$&$0.100^{+0.013}_{-0.016}$&\\
$m_{\nu,{\rm{sterile}}}^{\rm{eff}}$&$0.005$&$<0.701$&&$0.67$&$0.70\pm0.16$&&$0.62$&$0.68\pm0.15$&\\
$w$&$-1.030$&$-1.060^{+0.160}_{-0.110}$&&$-1.210$&$-1.200\pm0.100$&&$-1.167$&$-1.170^{+0.105}_{-0.096}$&\\
$N_{\rm eff}$&$3.44$&$3.72^{+0.31}_{-0.45}$&&$3.33$&$3.56^{+0.15}_{-0.33}$&&$3.80$&$3.81^{+0.28}_{-0.35}$&\\
$n_s$&$0.975$&$0.984^{+0.014}_{-0.017}$&&$0.962$&$0.974^{+0.011}_{-0.017}$&&$0.986$&$0.987^{+0.014}_{-0.013}$&\\
${\rm{ln}}(10^{10}A_s)$&$3.111$&$3.119^{+0.03}_{-0.034}$&&$3.087$&$3.110^{+0.028}_{-0.034}$&&$3.122$&$3.126^{+0.029}_{-0.034}$&\\
$r_{0.05}$&$0.007$&$<0.188$&&$0.006$&$<0.179$&&$0.203$&$0.186^{+0.034}_{-0.041}$&\\
\hline
$\Omega_\Lambda$&$0.709$&$0.702^{+0.017}_{-0.020}$&&$0.720$&$0.717^{+0.014}_{-0.013}$&&$0.716$&$0.713^{+0.014}_{-0.013}$&\\
$\Omega_m$&$0.291$&$0.298^{+0.020}_{-0.017}$&&$0.280$&$0.283^{+0.013}_{-0.014}$&&$0.284$&$0.287^{+0.013}_{-0.014}$&\\
$\sigma_8$&$0.851$&$0.816^{+0.038}_{-0.041}$&&$0.773$&$0.773\pm0.014$&&$0.776$&$0.772^{+0.014}_{-0.015}$&\\
$H_0$&$71.2$&$71.9^{+2.3}_{-3.2}$&&$73.0$&$73.7\pm2.0$&&$74.3$&$74.1\pm2.0$&\\
$r_{0.002}$&$0.006$&$<0.199$&&$0.006$&$<0.183$&&$0.210$&$0.193^{+0.038}_{-0.050}$&\\

\hline
$-\ln\mathcal{L}_{\rm{max}}$ &\multicolumn{2}{c}{4904.85} & & \multicolumn{2}{c}{4909.49} & & \multicolumn{2}{c}{4932.30} & \\
\hline
\end{tabular}
\caption{\label{tab4} Fitting results for the $w$CDM+$r$+$N_{\rm eff}$+$m_{\nu,{\rm sterile}}^{\rm eff}$ model. We quote $\pm 1\sigma$ errors, but for the parameters that cannot be well constrained, we quote the 95\% CL upper limits.}
\end{table}


 \end{appendices}



\begin{thebibliography}{99}

\bibitem{wmap5} 
  E.~Komatsu {\it et al.}  [WMAP Collaboration],
  ``Five-Year Wilkinson Microwave Anisotropy Probe (WMAP) Observations: Cosmological Interpretation,''
  Astrophys.\ J.\ Suppl.\  {\bf 180}, 330 (2009)
  [arXiv:0803.0547 [astro-ph]].

\bibitem{wmap7} 
  E.~Komatsu {\it et al.}  [WMAP Collaboration],
  ``Seven-Year Wilkinson Microwave Anisotropy Probe (WMAP) Observations: Cosmological Interpretation,''
  Astrophys.\ J.\ Suppl.\  {\bf 192}, 18 (2011)
  [arXiv:1001.4538 [astro-ph.CO]].


\bibitem{wmap9} 
  G.~Hinshaw {\it et al.}  [WMAP Collaboration],
  ``Nine-Year Wilkinson Microwave Anisotropy Probe (WMAP) Observations: Cosmological Parameter Results,''
  Astrophys.\ J.\ Suppl.\  {\bf 208}, 19 (2013)
  [arXiv:1212.5226 [astro-ph.CO]].
  
  
\bibitem{planck}
  P.~A.~R.~Ade {\it et al.}  [Planck Collaboration],
  ``Planck 2013 results. XVI. Cosmological parameters,''
  arXiv:1303.5076 [astro-ph.CO].
  
  
\bibitem{SZ} 
  P.~A.~R.~Ade {\it et al.}  [Planck Collaboration],
  arXiv:1303.5080 [astro-ph.CO].
  
  
\bibitem{shear} 
  J.~Benjamin, L.~Van Waerbeke, C.~Heymans, M.~Kilbinger, T.~Erben, H.~Hildebrandt, H.~Hoekstra and T.~D.~Kitching {\it et al.},
  arXiv:1212.3327 [astro-ph.CO].


\bibitem{snu3} 
  R.~A.~Battye and A.~Moss,
  ``Evidence for massive neutrinos from CMB and lensing observations,''
  Phys.\ Rev.\ Lett.\  {\bf 112}, 051303 (2014)
  [arXiv:1308.5870 [astro-ph.CO]].


  
  
\bibitem{h0} 
  A.~G.~Riess, L.~Macri, S.~Casertano, H.~Lampeitl, H.~C.~Ferguson, A.~V.~Filippenko, S.~W.~Jha and W.~Li {\it et al.},
  ``A 3\% Solution: Determination of the Hubble Constant with the Hubble Space Telescope and Wide Field Camera 3,''
  Astrophys.\ J.\  {\bf 730}, 119 (2011)
  [Erratum-ibid.\  {\bf 732}, 129 (2011)]
  [arXiv:1103.2976 [astro-ph.CO]].
 
  
\bibitem{lsnd} 
  A.~Aguilar-Arevalo {\it et al.}  [LSND Collaboration],
  ``Evidence for neutrino oscillations from the observation of anti-neutrino(electron) appearance in a anti-neutrino(muon) beam,''
  Phys.\ Rev.\ D {\bf 64}, 112007 (2001).


\bibitem{miniboone} 
  A.~A.~Aguilar-Arevalo {\it et al.}  [MiniBooNE Collaboration],
  ``Improved Search for $\bar \nu_\mu \rightarrow \bar \nu_e$ Oscillations in the MiniBooNE Experiment,''
  Phys.\ Rev.\ Lett.\  {\bf 110}, 161801 (2013).


\bibitem{reactor} 
  G.~Mention, M.~Fechner, T.~.Lasserre, T.~.A.~Mueller, D.~Lhuillier, M.~Cribier and A.~Letourneau,
  ``The Reactor Antineutrino Anomaly,''
  Phys.\ Rev.\ D {\bf 83}, 073006 (2011).



\bibitem{Giunti:2012tn} 
  C.~Giunti, M.~Laveder, Y.~F.~Li, Q.~Y.~Liu and H.~W.~Long,
  ``Update of Short-Baseline Electron Neutrino and Antineutrino Disappearance,''
  Phys.\ Rev.\ D {\bf 86}, 113014 (2012)
  [arXiv:1210.5715 [hep-ph]].


\bibitem{Giunti:2013aea} 
  C.~Giunti, M.~Laveder, Y.~F.~Li and H.~W.~Long,
  ``Pragmatic View of Short-Baseline Neutrino Oscillations,''
  Phys.\ Rev.\ D {\bf 88}, 073008 (2013)
  [arXiv:1308.5288 [hep-ph]].

\bibitem{snu1} 
  M.~Wyman, D.~H.~Rudd, R.~A.~Vanderveld and W.~Hu,
  ``$\nu\Lambda$CDM: Neutrinos help reconcile Planck with the Local Universe,''
  Phys.\ Rev.\ Lett.\  {\bf 112}, 051302 (2014)
  [arXiv:1307.7715 [astro-ph.CO]].
  
\bibitem{snu2} 
  J.~Hamann and J.~Hasenkamp,
  ``A new life for sterile neutrinos: resolving inconsistencies using hot dark matter,''
  JCAP {\bf 1310}, 044 (2013)
  [arXiv:1308.3255 [astro-ph.CO]].






\bibitem{bicep2}
  P.~A.~R.~Ade {\it et al.}  [BICEP2 Collaboration],
  ``Detection of B-Mode Polarization at Degree Angular Scales by BICEP2,''
  Phys.\ Rev.\ Lett.\  {\bf 112}, 241101 (2014)
  [arXiv:1403.3985 [astro-ph.CO]].
  
  
  
\bibitem{Liu:2014mpa} 
  H.~Liu, P.~Mertsch and S.~Sarkar,
  ``Fingerprints of Galactic Loop I on the Cosmic Microwave Background,''
  arXiv:1404.1899 [astro-ph.CO].

\bibitem{Harigaya:2014qza} 
  K.~Harigaya and T.~T.~Yanagida,
  ``Discovery of Large Scale Tensor Mode and Chaotic Inflation in Supergravity,''
  arXiv:1403.4729 [hep-ph].

\bibitem{Nakayama:2014koa} 
  K.~Nakayama and F.~Takahashi,
  ``Higgs Chaotic Inflation and the Primordial B-mode Polarization Discovered by BICEP2,''
  arXiv:1403.4132 [hep-ph].

\bibitem{Brandenberger:2014faa} 
  R.~H.~Brandenberger, A.~Nayeri and S.~P.~Patil,
  ``Closed String Thermodynamics and a Blue Tensor Spectrum,''
  arXiv:1403.4927 [astro-ph.CO].


\bibitem{Contaldi:2014zua} 
  C.~R.~Contaldi, M.~Peloso and L.~Sorbo,
  ``Suppressing the impact of a high tensor-to-scalar ratio on the temperature anisotropies,''
  arXiv:1403.4596 [astro-ph.CO].

\bibitem{Miranda:2014wga} 
  V.~'c.~Miranda, W.~Hu and P.~Adshead,
  ``Steps to Reconcile Inflationary Tensor and Scalar Spectra,''
  arXiv:1403.5231 [astro-ph.CO].


\bibitem{Gerbino:2014eqa} 
  M.~Gerbino, A.~Marchini, L.~Pagano, L.~Salvati, E.~Di Valentino and A.~Melchiorri,
  ``Blue Gravity Waves from BICEP2 ?,''
  arXiv:1403.5732 [astro-ph.CO].

\bibitem{McDonald:2014kia} 
  J.~McDonald,
  ``Negative Running of the Spectral Index, Hemispherical Asymmetry and Consistency of Planck with BICEP2,''
  arXiv:1403.6650 [astro-ph.CO].

\bibitem{Hazra:2014a} 
  D.~K.~Hazra, A.~Shafieloo, G.~F.~Smoot and A.~A.~Starobinsky,
  ``Ruling out the power-law form of the scalar primordial spectrum,''
  arXiv:1403.7786 [astro-ph.CO].


\bibitem{Hazra:2014b} 
  D.~K.~Hazra, A.~Shafieloo, G.~F.~Smoot and A.~A.~Starobinsky,
  ``Whipped inflation,''
  arXiv:1404.0360 [astro-ph.CO].


\bibitem{Kehagias:2014wza} 
  A.~Kehagias and A.~Riotto,
  ``Remarks about the Tensor Mode Detection by the BICEP2 Collaboration and the Super-Planckian Excursions of the Inflaton Field,''
  arXiv:1403.4811 [astro-ph.CO].

\bibitem{Lyth:2014yya} 
  D.~H.~Lyth,
  ``BICEP2, the curvature perturbation and supersymmetry,''
  arXiv:1403.7323 [hep-ph].

\bibitem{Bonvin:2014xia} 
  C.~Bonvin, R.~Durrer and R.~Maartens,
  ``Can primordial magnetic fields be the origin of the BICEP2 data?,''
  arXiv:1403.6768 [astro-ph.CO].

\bibitem{Lizarraga:2014eaa} 
  J.~Lizarraga, J.~Urrestilla, D.~Daverio, M.~Hindmarsh, M.~Kunz and A.~R.~Liddle,
  ``Can topological defects mimic the BICEP2 B-mode signal?,''
  arXiv:1403.4924 [astro-ph.CO].

\bibitem{Moss:2014cra} 
  A.~Moss and L.~Pogosian,
  ``Did BICEP2 see vector modes? First B-mode constraints on cosmic defects,''
  arXiv:1403.6105 [astro-ph.CO].


\bibitem{Chluba:2014uba} 
  J.~Chluba, L.~Dai, D.~Jeong, M.~Kamionkowski and A.~Yoho,
  ``Linking the BICEP2 result and the hemispherical power asymmetry through spatial variation of $r$,''
  arXiv:1404.2798 [astro-ph.CO].
  
  
\bibitem{zx14} 
  J.~-F.~Zhang, Y.~-H.~Li and X.~Zhang,
  ``Sterile neutrinos help reconcile the observational results of primordial gravitational waves from Planck and BICEP2,''
  arXiv:1403.7028 [astro-ph.CO].

\bibitem{WHu14} 
  C.~Dvorkin, M.~Wyman, D.~H.~Rudd and W.~Hu,
  ``Neutrinos help reconcile Planck measurements with both Early and Local Universe,''
  arXiv:1403.8049 [astro-ph.CO].

 
\bibitem{sterile2} 
  J.~-F.~Zhang, Y.~-H.~Li and X.~Zhang,
  ``Cosmological constraints on neutrinos after BICEP2,''
  Eur.\ Phys.\ J.\ C {\bf 74}, 2954 (2014)
  [arXiv:1404.3598 [astro-ph.CO]].
  
  
\bibitem{sterile3} 
  Y.~-H.~Li, J.~-F.~Zhang and X.~Zhang,
  ``Tilt of primordial gravitational wave spectrum in a universe with sterile neutrinos,''
  Sci.\ China Phys.\ Mech.\ Astron.\  {\bf 57}, 1455 (2014)
  [arXiv:1405.0570 [astro-ph.CO]].
  
  
\bibitem{nus14a} 
  M.~Archidiacono, N.~Fornengo, S.~Gariazzo, C.~Giunti, S.~Hannestad and M.~Laveder,
  ``Light sterile neutrinos after BICEP-2,''
  JCAP {\bf 1406}, 031 (2014)
  [arXiv:1404.1794 [astro-ph.CO]].


\bibitem{nus14b} 
  J.~Bergstršm, M.~C.~Gonzalez-Garcia, V.~Niro and J.~Salvado,
  ``Statistical tests of sterile neutrinos using cosmology and short-baseline data,''
  arXiv:1407.3806 [hep-ph].


\bibitem{hde} 
  M.~Li, X.~-D.~Li, Y.~-Z.~Ma, X.~Zhang and Z.~Zhang,
  ``Planck Constraints on Holographic Dark Energy,''
  JCAP {\bf 1309}, 021 (2013)
  [arXiv:1305.5302 [astro-ph.CO]].
  
  
\bibitem{Lewis:2002ah} 
  A.~Lewis and S.~Bridle,
  ``Cosmological parameters from CMB and other data: A Monte Carlo approach,''
  Phys.\ Rev.\ D {\bf 66}, 103511 (2002)
  [astro-ph/0205436].


\bibitem{boss} 
  L.~Anderson {\it et al.}  [BOSS Collaboration],
  ``The clustering of galaxies in the SDSS-III Baryon Oscillation Spectroscopic Survey: Baryon Acoustic Oscillations in the Data Release 10 and 11 galaxy samples,''
  arXiv:1312.4877 [astro-ph.CO].


\bibitem{6df} 
  F.~Beutler, C.~Blake, M.~Colless, D.~H.~Jones, L.~Staveley-Smith, L.~Campbell, Q.~Parker and W.~Saunders {\it et al.},
  ``The 6dF Galaxy Survey: Baryon Acoustic Oscillations and the Local Hubble Constant,''
  Mon.\ Not.\ Roy.\ Astron.\ Soc.\  {\bf 416}, 3017 (2011)
  [arXiv:1106.3366 [astro-ph.CO]].


\bibitem{sdss7} 
  W.~J.~Percival {\it et al.}  [SDSS Collaboration],
  ``Baryon Acoustic Oscillations in the Sloan Digital Sky Survey Data Release 7 Galaxy Sample,''
  Mon.\ Not.\ Roy.\ Astron.\ Soc.\  {\bf 401}, 2148 (2010)
  [arXiv:0907.1660 [astro-ph.CO]].



\bibitem{wigglez} 
  C.~Blake, E.~Kazin, F.~Beutler, T.~Davis, D.~Parkinson, S.~Brough, M.~Colless and C.~Contreras {\it et al.},
  ``The WiggleZ Dark Energy Survey: mapping the distance-redshift relation with baryon acoustic oscillations,''
  Mon.\ Not.\ Roy.\ Astron.\ Soc.\  {\bf 418}, 1707 (2011)
  [arXiv:1108.2635 [astro-ph.CO]].
 
  
\bibitem{Tinker2008} 
  J.~L.~Tinker, A.~V.~Kravtsov, A.~Klypin, K.~Abazajian, M.~S.~Warren, G.~Yepes, S.~Gottlober and D.~E.~Holz,
  ``Toward a halo mass function for precision cosmology: The Limits of universality,''
  Astrophys.\ J.\  {\bf 688}, 709 (2008)
  [arXiv:0803.2706 [astro-ph]].


\bibitem{Watson2013}
W. A. Watson, I. T. Iliev, A. D'Aloisio, A. Knebe, P. R. Shapiro and G. Yepes, 
``The halo mass function through the cosmic ages,''
Mon.\ Not.\ Roy.\ Astron.\ Soc.\  {\bf 433}, 1230 (2013) 
[arXiv:1212.0095 [astro-ph.CO]].

  
\bibitem{cmblensing} 
  P.~A.~R.~Ade {\it et al.}  [Planck Collaboration],
  ``Planck 2013 results. XVII. Gravitational lensing by large-scale structure,''
  arXiv:1303.5077 [astro-ph.CO].
  
  
  
  
 
  
  
\end{thebibliography}
\end{document}